\documentclass[a4paper,11pt]{article}
\usepackage[dvips,a4paper,text={16cm,24cm},centering]{geometry}
\usepackage[latin1]{inputenc}
\usepackage[T1]{fontenc}
\usepackage{amssymb}
\usepackage{amsmath}
\usepackage{graphicx}
\usepackage{pstricks}
\usepackage{epic}

\numberwithin{equation}{section}

\newcommand{\ZZ}{\mathbb{Z}}

\newcommand{\be}{\begin{equation}}
\newcommand{\ee}{\end{equation}}
\newcommand{\bea}{\begin{eqnarray}}
\newcommand{\eea}{\end{eqnarray}}
\newcommand{\ud}{\mathrm{d}}
\newcommand{\G}{\left}
\newcommand{\D}{\right}
\newcommand{\p}{\partial}
\newcommand{\w}{\wedge}
\newcommand{\ie}{\emph{i.e. }}
\newcommand{\cA}{\mathcal{A}}
\newcommand{\cH}{\mathcal{H}}
\newcommand{\cV}{\mathcal{V}}
\newcommand{\cG}{\mathcal{G}}
\newcommand{\cL}{\mathcal{L}}

\newcommand{\cP}{\mathcal{P}}
\newcommand{\cQ}{\mathcal{Q}}

\newcommand{\te}{\tilde{e}}
\newcommand{\tf}{\tilde{f}}
\newcommand{\tk}{\tilde{k}}
\newcommand{\0}{{(0)}}
\newcommand{\1}{{(1)}}
\newcommand{\2}{{(2)}}
\newcommand{\3}{{(3)}}
\newcommand{\4}{{(4)}}
\newcommand{\5}{{(5)}}

\newcommand{\bpsi}{\overline{\psi}}
\newcommand{\hpsi}{\hat{\psi}}
\newcommand{\bhpsi}{\overline{\hat{\psi}}}

\newcommand{\he}{\hat{e}}
\newcommand{\hgamma}{\hat{\gamma}}

\newcommand{\Sh}[1]{#1\hskip-9.5pt \diagup}

\def\a{\alpha}
\def\b{\beta}
\def\g{\gamma}
\def\d{\delta}
\def\h{\eta}

\def\m{\mu}
\def\n{\nu}
\def\r{\rho}
\def\o{\omega}
\def\O{\Omega}
\def\s{\sigma}
\def\S{\Sigma}

\def\P{\Pi}

\def\ft#1#2{{\textstyle{{\scriptstyle #1}\over {\scriptstyle #2}}}}
\def\oneone{\rlap 1\mkern4mu{\rm l}}
\def\sst#1{{\scriptscriptstyle #1}}
\def\Dm{{{D_{\sst{max}}}}}
\def\cF{{\cal F}}
\def\cA{{\cal A}}
\def\nn{\nnnumber}
\def\no{\nonumber}

\def\td{\tilde}
\def\hg{\hat{\gamma}}
\def\e{\epsilon}

\def\fft#1#2{{#1 \over #2}}
\def\nn{\nonumber}
\def\R{\rlap{\rm I}\mkern3mu{\rm R}}
\def\hA{\hat{\cal A}}
\def\cQ{ { \cal Q } }
\begin{document}

\begin{flushright}
ULB-TH/05-13\\
May 2005\\
\vspace*{1cm}
\end{flushright}

\begin{center}
\begin{Large}
\textbf{Hidden Symmetries and Dirac Fermions}
\end{Large}

\vspace{7mm} {\bf Sophie de Buyl\footnote{Aspirant du Fonds
National de la Recherche Scientifique, Belgique}, Marc
Henneaux\footnote{Also at CECS, Valdivia, Chile}, Louis Paulot}

\vspace{5mm}
Physique théorique et mathématique, Université libre de Bruxelles\\
and\\
International Solvay Institutes\\
Campus Plaine C.P.~231, B--1050 Bruxelles, Belgium \vspace{3mm}

\end{center}

\vspace{3mm} \hrule
\begin{abstract}

In this paper, two things are done.  First, we analyze the
compatibility of Dirac fermions with the hidden duality symmetries
which appear in the toroidal compactification of gravitational
theories down to three spacetime dimensions.  We show that the Pauli
couplings to the $p$-forms can be adjusted, for all simple (split)
groups, so that the fermions transform in a representation of the
maximal compact subgroup of the duality group $G$ in three
dimensions. Second, we investigate how the Dirac fermions fit in the
conjectured hidden overextended symmetry $G^{++}$.  We show
compatibility with this symmetry up to the same level as in the pure
bosonic case.  We also investigate the BKL behaviour of the
Einstein-Dirac-$p$-form systems and provide a group theoretical
interpretation of the Belinskii-Khalatnikov result that the Dirac
field removes chaos.

\end{abstract}
\hrule

\vspace{3mm}

\section{Introduction}
The emergence of unexpected (``hidden") symmetries in the toroidal
dimensional reduction of gravitational theories to lower
dimensions is a fascinating discovery whose full implications are
to a large extent still mysterious \cite{Geroch,CJ0}. This result,
which underlies $U$-duality \cite{Pio}, appears quite clearly when
one dimensionally reduces down to $3$ spacetime dimensions since
the Lagrangian can then be rewritten as the coupled
Einstein-scalar Lagrangian, where the scalars parametrize the
symmetric space $G/K(G)$. Here, $G$ is the ``hidden" symmetry
group and $K(G)$ its maximal compact subgroup (see
\cite{Cremmer:1999du} for a systematic study as well as the
earlier work \cite{Breitenlohner:1987dg}). It has been argued that
this $G$-symmetry signals a much bigger, infinite-dimensional,
symmetry, which would be the overextension $G^{++}$
\cite{JuliaInf,Nicolai,DH1,HJ,DHN}, or even the very extended
extension $G^{+++}$, of $G$ \cite{West,EH} (or perhaps a Borcherds
superalgebra related to it \cite{Henry-LabordereJP}).

One intriguing feature of the hidden symmetries is the fact that
when the coupled Einstein-$G/K(G)$ system is the bosonic sector of
a supergravity theory, then, important properties of
supergravities which are usually derived on the grounds of
supersymmetry may alternatively be obtained by invoking the hidden
symmetries. This is for instance the case of the Chern-Simons term
and of the precise value of its coefficient in eleven dimensional
supergravity, which is required by supersymmetry
\cite{Cremmer:1978km}, but which also follows from the $E$ ($E_8$
or $E_{10}$) symmetry of the Lagrangian \cite{CJ0,DHN}. Quite
generally, the spacetime dimension $11$ is quite special for the
Einstein-$3$-form system, both from the point of view of
supersymmetry and from the existence of hidden symmetries. Another
example will be provided below (subsection \ref{e8}). One might
thus be inclined to think that there is a deep connection between
hidden symmetries and supergravity. However, hidden symmetries
exist even for bosonic theories that are not the bosonic sectors
of supersymmetric theories. For this reason, they appear to have a
wider scope.

In order to further elucidate hidden symmetries, we have
investigated how fermions enter the picture. Although the
supersymmetric case is most likely ultimately the most
interesting, we have considered here only spin-$1/2$ fermions, for
two reasons. First, this case is technically simpler. Second, in
the light of the above comments, we want to deepen the
understanding of the connection ---~or the absence of connection~---
between hidden symmetries and supersymmetry.

The Einstein-($p$-form)-Dirac system by itself is not
supersymmetric and yet we find that the Dirac fermions are
compatible with the $G$-symmetry, for all (split) real simple Lie
groups. Indeed, one may arrange for the fermions to form
representations of the compact subgroup $K(G)$. This is automatic
for the pure Einstein-Dirac system. When $p$-forms are present,
the hidden symmetry invariance requirement fixes the Pauli
couplings of the Dirac fermions with the $p$-forms, a feature
familiar from supersymmetry. In particular, $E_8$-invariance of
the coupling of a Dirac fermion to the Einstein-$3$-form system
reproduces the supersymmetric covariant Dirac operator of
$11$-dimensional supergravity \cite{Cremmer:1978km,Julia:1986qq}.
A similar feature holds for ${\cal N}=1$ supergravity in $5$
dimensions \cite{ChamNic}.  Thus, we see again that hidden
symmetries of gravitational theories appear to have a wider scope
than supersymmetry but yet, have the puzzling feature of
predicting similar structures when supersymmetry is available.

We formulate the theories both in $3$ spacetime dimensions, where
the symmetries are manifest, and in the maximum oxidation dimension,
where the Lagrangian is simpler. To a large extent, one may thus
view our paper as an extension of the oxidation analysis
\cite{JuliaInf,J2,Cremmer:1999du,HJ,Henry-LabordereJP,Keur,dBHJP} to
include Dirac fermions. Indeed the symmetric Lagrangian is known in
$3$ dimensions and one may ask the question of how high it oxidizes.
It turns out that in most cases, the Dirac fermions do not bring new
obstructions to oxidation in addition to the ones found in the
bosonic sectors. If the bosonic Lagrangian lifts up to $n$ dimensions,
then the coupled bosonic-Dirac Lagrangian (with the Dirac fields
transforming in appropriate representations of the maximal compact
subgroup $K(G)$) also lifts up to $n$ dimensions. This absence of
new obstructions coming from the fermions is in line with the
results of Keurentjes \cite{Keur3}, who has shown that the topology
of the compact subgroup $K(G)$ is always appropriate to allow for
fermions in higher dimensions when the bosonic sector can be
oxidized.

We then investigate how the fermions fit in the conjectured
infinite-dimensional symmetry $G^{++}$ and find indications that
the fermions form representations of $K(G^{++})$) up to the level
where the matching works for the bosonic sector. We study next the
BKL limit \cite{BKL,DHN2} of the systems with fermions. We extend
to all dimensions the results of \cite{BK}, where it was found
that the inclusion of Dirac spinors (with a non-vanishing
expectation value for fermionic currents) eliminates chaos in four
dimensions. Our analysis provides furthermore a group theoretical
interpretation of this result: elimination of chaos follows from
the fact that the geodesic motion on the symmetric space
$G^{++}/K(G^{++})$, which is lightlike in the pure bosonic case
\cite{DHN2}, becomes timelike when spin-$1/2$ fields are included
(and their currents acquire non-vanishing values) --- the mass term
being given by the Casimir of the maximal compact subgroup
$K(G^{++})$ in the fermionic representation.

Our paper is organized as follows. In the next section, we collect
conventions and notations. We then consider the dimensional
reduction to three dimensions of the pure Einstein-Dirac system in
$D$ spacetime dimensions and show that the fermions transform in
the spinorial representation of the maximal compact subgroup
$SO(n+1)$ in three dimensions, as they should (section
\ref{gravity}). The $SO(n,n)$ case is treated next, by relying on
the pure gravitational case. The maximal compact subgroup is now
$SO(n) \times SO(n)$. We show that one can choose the Pauli
couplings so that the fermions transform in a representation of
$SO(n) \times SO(n)$ (in fact, one can adjust the Pauli couplings
so that different representations arise) (section \ref{dncase}).
In section \ref{e8case}, we turn to the $E_n$-family. We show that
again, the Pauli couplings can be adjusted so that the spin-$1/2$
fields transform in a representation of $SO(16)$, $SU(8)$ or
$Sp(4)$ when one oxidizes the theory along the standard lines. 
We then show that the
$G_2$-case admits also covariant fermions in 5 dimensions (section
\ref{g2case}) and treat next all the other non simply laced groups
from their embeddings in simply laced ones (section
\ref{nonsimplylaced}).

In section \ref{G++}, we show that the Dirac fields fit (up to the
same level as the bosons) into the conjectured $G^{++}$ symmetry,
by considering the coupling of Dirac fermions to the $(1+0)$ non
linear sigma model of \cite{DHN}. In section \ref{BBKKLL}, we
analyze the BKL limit and argue that chaos is eliminated by the
Dirac field because the Casimir of the $K(G^{++})$ currents in the
fermionic representation provides a mass term for the geodesic
motion on the symmetric space $G^{++}/K(G^{++})$. Finally, we
close our paper with some conclusions and technical appendices.

In the analysis of the models, we rely very much on the papers
\cite{Cremmer:1997ct,Cremmer:1998px,Cremmer:1999du}, where the
maximally oxidized theories have been worked out in detail and
where the patterns of dimensional reduction that we shall need
have been established.

\section{Conventions}
\label{conventions}
\subsection{Chevalley-Serre presentation and Cartan-Chevalley involution}
We adopt the standard Chevalley-Serre presentation of the Lie
algebras in terms of $3r$ generators $\{h_i,e_i,f_i\}$ ($i= 1,
\cdots, r$ with $r$ equal to the rank) as given for instance in
\cite{Kac}, except that we take the ``negative" generators $f_i$
with the opposite sign (with respect to \cite{Kac}). The only
relation that is modified is $[e_i, f_j] = - \delta_{ij} \, h_i $.
The other relations are unchanged, namely, $[h_i, h_j] = 0$,
$[h_i, e_j] = A_{ij} e_j$, $[h_i, f_j] = -A_{ij} f_j$ (where
$A_{ij}$ is the Cartan matrix) and $ad_{e_i}^{1-A_{ij}} e_j = 0$,
$ad_{f_i}^{1-A_{ij}} f_j = 0$.

The sign convention for $f_i$ simplifies somewhat the form of the
generators of the maximal compact subalgebra. The Cartan-Chevalley
involution reads $\tau(h_i) = -h_i$, $\tau(e_i) = f_i$, $\tau(f_i)
= e_i$ and extends to the higher height root vectors as
$\tau(e_\alpha) = f_{\alpha}$, $\tau(f_{\alpha}) = e_{\alpha}$
where the $e_\alpha$'s are given by multi-commutators of the
$e_i$'s and the $f_{\alpha}$'s are given by multi-commutators of
the $f_i$'s in the same order (e.g., if $e_{r+1} = [e_1, e_2]$,
then $f_{r+1} = [f_1,f_2]$). With the opposite sign convention for
$f_i$, one has $\tau(e_i) = - f_i$ and the less uniform rule
$\tau(e_\alpha) = (-1)^{ht(\alpha)} f_\alpha$. A basis of the
maximal compact subalgebra is given by $k_\alpha = e_{\alpha} +
f_{\alpha}$. It is convenient to define $\cG^T = - \tau(\cG)$ for
any Lie algebra element $\cG$.

We shall also be interested in infinite-dimensional (Lorentzian)
Kac-Moody algebras, for which we adopt the same conventions. In
that case, the root space associated with imaginary roots might be
degenerate and we therefore add an index $s$ to account for the
degeneracy, $e_{\alpha} \rightarrow e_{\alpha,s}$, $f_{\alpha}
\rightarrow f_{\alpha,s}$, etc.

We shall exclusively deal in this paper with the split real forms
of the Lie algebras, defined as above in terms of the same
Chevalley-Serre presentation but with coefficients that are
restricted to be real numbers. Remarks on the non-split case are
given in the conclusions.

\subsection{Invariant bilinear form}
The invariant bilinear form on the Lie algebra is given for the
Chevalley-Serre generators by \be K(h_i,h_j) = \frac{2 \,
A_{ji}}{(\alpha_i \vert \alpha_i)} = \frac{2 \, A_{ij}}{(\alpha_j
\vert \alpha_j)}, \;\; K(h_i,e_j)= K(h_i, f_j)=0, \; \; K(e_i,f_j)
= - \frac{2 \delta_{ij}}{(\alpha_i \vert \alpha_i)} \ee and is
extended to the full algebra by using the invariance relation
$K(x,[y,z]) = K([x,y],z)$. Here, the $\alpha_i$'s are the simple
roots. The induced bilinear form in root space is denoted by $( \cdot
\vert \cdot )$ and given by $(\alpha_i \vert \alpha_j) = \frac{2
A_{ij}}{(\alpha_i \vert \alpha_i)}$. The numbers $(\alpha_i \vert
\alpha_i)$ are such that the product $A_{ij} \, (\alpha_i \vert
\alpha_i)$ is symmetric and they are normalized so that the
longest roots have squared length equal to $2$.   One gets \be
K(h_i,e_\alpha)= K(h_i, f_\alpha) = 0 = K(e_\alpha, e_\beta) =
K(f_\alpha, f_\beta), \; \; K(e_\alpha, f_\beta) = - N_\a
\delta_{\alpha \beta}, \label{norma} \ee where the coefficient
$N_\a$ in front of $\delta_{\alpha \beta}$ in the last relation
depends on the Cartan matrix (and on the precise normalization of
the root vectors corresponding to higher roots -- e.g., $ N_\a =
\frac{2}{(\a \vert \a)}$ in the Cartan-Weyl-Chevalley basis). In a
representation $T$, the bilinear form $K(x,y)$ is proportional to
the trace $\mathrm{Tr} (T(x) T(y))$.

In the finite-dimensional case, the Cartan subalgebra $\cH$ is an
Euclidean vector space. As in \cite{Cremmer:1997ct}, we shall
sometimes find it convenient to use then an orthogonal basis
$\{H_i\}$ of the Cartan subalgebra such that \be K(H_i, H_j) = 2
\d_{ij} . \ee It follows that $K(\vec{a}, \vec{b}) = 2 \vec{a} .
\vec{b}$ for $\vec{a}$, $\vec{b}$ in $\cH$ ($\vec{a} = \sum_i a^i
H_i$, $\vec{b} = \sum_i b^i H_i$) and $(\vec{\a} \vert \vec{\b}) =
\frac{1}{2} \vec{\a} . \vec{\b}$ for $\vec{\a}$, $\vec{\b}$ in the
dual space. Here, $\vec{a} . \vec{b} = \sum_i a^i b^i$ and
$\vec{\a} . \vec{\b} = \sum_i \a_i \b_i$ (and $\a_i, \b_i$
components of $\vec{\a}$, $\vec{\b}$ in the dual basis).

\subsection{$K(G)$-connection}
We parametrize the coset space $G/K(G)$ (with the gauge subgroup
$K(G)$ acting from the left) by taking the group elements $\cV$ in
the upper-triangular ``Borel gauge". The differential \be \cG =
\ud \cV \cV^{-1} \rlap{} \ee is in the Borel algebra, i.e., is a
linear combination of the $h_i$'s and the $e_\alpha$'s. It is
invariant under right multiplication. One defines $\cP$ as its
symmetric part and $\cQ$ as its antisymmetric part, \be
\cP=\frac{1}{2}(\cG+\cG^T), \; \; \; \; \; \;
\cQ=\frac{1}{2}(\cG-\cG^T) \, . \ee $\cQ$ is in the compact
subalgebra. Under a gauge transformation $\cP$ is covariant
whereas the antisymmetric part $\cQ$ transforms as a gauge
connection, \be \cV \ \longrightarrow \ H \cV \rlap{\ ,} \; \; \;
\; \; \;  \cP \ \longrightarrow \ H \cP H^{-1}\rlap{\ ,} \; \; \;
\; \; \; \cQ \ \longrightarrow \ \ud H H^{-1}+ H \cQ H^{-1}  \ee
(with $H \in K(G)$).

One may parametrize $\cV$ as $\cV = \cV_1 \, \cV_2 $ where $\cV_1$
is in the Cartan torus $\cV_1 = e^{\frac{1}{2} \phi^i H_i}$ and
$\cV_2$ is in the nilpotent subgroup generated by the
$e_\alpha$'s. In terms of this parametrization, one finds \be \cG
= \frac{1}{2} \, \ud \phi^i H_i + \sum_{\alpha \in \Delta_+}
e^{\frac{1}{2} \vec{\a} . \vec{\phi}} \cF_\alpha e_\alpha
\label{formofG}\ee where $\Delta_+$ is the set of positive roots.
Here, the one-forms $\cF_\alpha$ are defined through \be \ud \cV_2
\cV_2^{-1} = \sum_{\alpha \in \Delta_+} \cF_\alpha e_\alpha. \ee
In the infinite-dimensional case, there is also a sum over the
multiplicity index. Thus we get \be \cP = \frac{1}{2}\, \ud \phi^i
H_i + \frac{1}{2} \sum_{\alpha \in \Delta_+} e^{\frac{1}{2}
\vec{\a} . \vec{\phi}} \cF_\alpha \left(e_\alpha- f_\alpha \right)
\label{formofP}\ee and \be \cQ = \frac{1}{2}\sum_{\alpha \in
\Delta_+} e^{\frac{1}{2} \vec{\a} . \vec{\phi}} \cF_\alpha \,
k_\alpha  \equiv \sum_{\alpha \in \Delta_+} \cQ_{(\a)} \, k_\a
\label{formofQ}\ee with \be \cQ_{ (\a)} = \frac{1}{2}e^{\frac{1}{2}
\vec{\a} . \vec{\phi}} \cF_{ \alpha}. \ee We see therefore that
the one-forms $(1/2) e^{\frac{1}{2} \vec{\a} . \vec{\phi}}
\cF_{\alpha }$ appear as the components of the connection one-form
of the compact subgroup $K(G)$ in the basis of the $k_\alpha$'s.

\subsection{Lagrangian for coset models}
The Lagrangian for the coset model $G/K(G)$ reads \be \cL_{G/K(G)}
= - K\G( \cP ,\w
* \cP \D) \ee If we expand the Lagrangian according to
(\ref{formofP}), we get \be {\cal L}_{G/K(G)} = - \ft12
{*d\vec{\phi}}\wedge d\vec{\phi} - \frac{1}{2}\sum_{\alpha \in
\Delta_+} N_\a \, e^{ \vec{\a} . \vec{\phi}}\, {*\cF_\alpha}\wedge
\cF_\alpha \label{generalform}\ee where the factor $N_\a$ is
defined in (\ref{norma}).

The coupling of a field $\psi$ transforming in a representation
$J$ of the ``unbroken" subgroup $K(G)$ is straightforward. One
replaces ordinary derivatives $\partial_\m$ by covariant
derivatives $D_\m$ where \be D_\m \psi = \partial_\m \psi-
\sum_{\alpha \in \Delta_+}\cQ_{\mu  (\alpha)} J_{\alpha} \psi
\label{cova}\ee with  $\cQ_{(\alpha)} = \cQ_{\m (\alpha) }dx^\m $.  In
(\ref{cova}), $J_{\alpha}$ are the generators of the representation
$J$ of $K(G)$ in which $\psi$ transforms, $J_{\alpha} = J(k_\alpha)$
(the generators $J_{\alpha}$ obeys the same commutation relations as
$k_\alpha$). This guarantees $K(G)$ -- and hence $G$ --
invariance. The three-dimensional Dirac Lagrangian is thus (in
flat space) \be \bar{\psi} \gamma^\m \left( \partial_\m
-\sum_{\alpha \in \Delta_+}\cQ_{\mu  (\alpha)} J_{\alpha} \right)
\psi \ee

\subsection{Dimensional reduction of metric and exterior forms}
{}For the purpose of being self-contained, we recall the general
formulas for dimensional reduction on a torus of the metric and a
$(p-1)$-form potential. We adhere to the conventions and notations
of \cite{Cremmer:1999du}, which we follow without change. We
consider reduction down to three spacetime dimensions.

The metric is reduced as \be ds_D^2 = e^{\vec{s}. \vec{\phi}}
ds^2_3 + \sum_{i=1}^n e^{2 \vec{\gamma}_i . \vec{\phi}} (h^i)^2
\label{metricreduction}\ee where the one-forms $h^i$ are given by
\be h^i = dz^i + \cA^i_{(0)j} \, dz^j + \cA^i_{(1)} =
\tilde{\gamma}^i_{\; j} (dz^j + \hat{\cA}^j_{(1)}) \ee with
$\tilde{\gamma}^i_{\; j} \equiv (\gamma^{-1})^i{}_j \equiv
\delta^i_j + \cA^i_{(0)j}$ and $\cA^i_{(0)j}$ non vanishing only
for $i<j$. The vector $\vec{\phi}$ collects the dilatons. The
$z^i$'s are internal coordinates on the torus, while
$\cA^i_{(0)j}$ and $\cA^i_{(1)}$ are respectively scalars and
$1$-forms in three dimensions. Furthermore, $\vec{s} = (s_1,
\ldots, s_n)$ and $\vec{\gamma}_i = \frac{1}{2}
(s_1,\ldots,s_{i-1},(2+i-D)s_i,0,\ldots,0)$ where $$s_i =
\sqrt{\frac{2}{(D-1-i)(D-2-i)}}$$.

A $(p-1)$-form potential $A_{p-1}$ decomposes as a sum of
$1$-forms and of scalars in three dimensions, \be A_{p-1} =
A_{(1)i_1 \dots i_{p-2}} dz^{i_1} \wedge \cdots \wedge
dz^{i_{p-2}} + A_{(0)i_1 \dots i_{p-1}} dz^{i_1} \wedge \cdots
\wedge dz^{i_{p-1}}\ee (the $2$-form component carries no degree
of freedom and is dropped).

\section{Dimensional reduction of the Einstein-Dirac System}
\label{gravity}

\subsection{Reduction of gravity}

We start with the simplest case, namely, that of the coupled
Einstein-Dirac system without extra fields. The bosonic sector
reduces to pure gravity. To show that the Dirac field is
compatible with the hidden symmetry is rather direct in this case.

Upon dimensional reduction down to $d=3$, gravity in $D=3+n$ gets
a symmetry group $SL(n+1)$, beyond the $SL(n)$
symmetry of the reduced dimensions. Moreover, in three dimensions
the scalars describe a $SL(n+1)/SO(n+1)$
$\sigma$-model. Following (\ref{metricreduction}), we parametrize
the $D=3+n$ vielbein in a triangular gauge as \be e = \G(
\begin{array}{cc} e^{\frac{1}{2} \vec{s}.\vec{\phi}} \hat{e}_\mu{}^m
&
e^{\vec{\gamma_i}.\vec{\phi}} \cA_\1{}_\mu{}^i \\
0 & M_i{}^j
\end{array} \D)
\label{vb} \ee with $\mu,m = 0,D-2,D-1$ and $i,j = 1..n$. We
choose the non-compactified dimensions to be $0$, $D-2$ and $D-1$
so that indices remain simple in formulas . The triad in three
spacetime dimensions is $\hat{e}_\mu{}^m$. We denote by $M$ the
upper-triangular matrix \be M_i{}^j =
e^{\vec{\gamma_i}.\vec{\phi}} (\delta_i^j + \cA_\0{}^i{}_j) =
e^{\vec{\gamma_i}.\vec{\phi}} (\gamma^{-1})^i{}_j \ee One can
check that $\det(M) = e^{-\frac{1}{2}\vec{s}.\vec{\phi}}$.

After dualizing the Kaluza-Klein vectors $\cA_\1{}^i$ into scalars
$\chi_j$ as \be e^{\vec{b}_i.\vec{\phi}} * \! \left(\td\g^i{}_j d
(\g^j{}_m {\cA}_\1^m )\right) = \gamma^j{}_i \G( \ud \chi_j \D)
\label{dual-a1} \ee one can form the upper-triangular $(n+1)
\times (n+1)$ matrix \be \cV^{-1} = \G(
\begin{array}{cc}
M_i{}^j & \chi_i e^{\frac{1}{2}\vec{s}.\vec{\phi}} \\
0 & e^{\frac{1}{2} \vec{s}.\vec{\phi}}
\end{array} \D)
\label{V-An} \ee which parametrizes a
$SL(n+1)/SO(n+1)$ symmetric space. With this
parametrization, the three-dimensional reduced Lagrangian becomes
\be \cL_E^\3 = \hat{R} *1 - \frac{1}{2} \mathrm{Tr}\G( \cP \w *
\cP \D) \ee where one finds explicitly from (\ref{V-An}) \be \cG =
\ud \cV \cV^{-1}= - \cV (\ud \cV^{-1}) = - \G(
\begin{array}{cc}
M^{-1} \ud M & M^{-1} \ud \chi e^{\frac{1}{2}\vec{s}.\vec{\phi}} \\
0 & \frac{1}{2} \vec{s}.\ud \vec{\phi}
\end{array} \D)
\ee and \bea \cP &=& - \G(
\begin{array}{cc} \frac{1}{2} \G( M^{-1} \ud M + (M^{-1} \ud M)^T
\D)
& \frac{1}{2} M^{-1} \ud \chi e^{\frac{1}{2}\vec{s}.\vec{\phi}} \\
\frac{1}{2} \G(M^{-1} \ud \chi e^{\frac{1}{2}\vec{s}.\vec{\phi}}
\D)^T & \frac{1}{2} \vec{s}.\ud \vec{\phi}
\end{array} \D) \\
\cQ &=& - \G( \begin{array}{cc} \frac{1}{2} \G( M^{-1} \ud M -
(M^{-1} \ud M)^T \D)
& \frac{1}{2} M^{-1} \ud \chi e^{\frac{1}{2}\vec{s}.\vec{\phi}} \\
-\frac{1}{2} \G(M^{-1} \ud \chi e^{\frac{1}{2}\vec{s}.\vec{\phi}}
\D)^T & 0
\end{array} \D)
\label{Q-An} \rlap{\ .} \eea

\subsection{Adding spinors}

We now couple a Dirac spinor to gravity in $D=3+n$ dimensions: \be
\cL = \cL_E + \cL_D \ee with $\cL_E$ the Einstein Lagrangian and
$\cL_D$ the Dirac Lagrangian, \be \cL_D = e \bpsi \Sh{D} \psi = e
\bpsi \gamma^\S \G( \p_\S - \frac{1}{4} \omega_{\S,AB} \gamma^{AB}
\D) \psi \ee $e$ is the determinant of the vielbein and $\S, A =
0,... ,\Dm - 1$. The indices $A$, $B$, ... are internal indices
and $\Sigma$, $\Omega$ are spacetime indices, while $\omega$ is
the spin connection, which can be computed from the vielbein: \bea
\omega_{\S,AB} &=& \frac{1}{2} e_A{}^\O (\p_\O e_\S{}_B - \p_\S
e_\O{}_B) - \frac{1}{2} e_B{}^\O (\p_\O e_\S{}_A - \p_\S e_\O{}_A
)  \nonumber \\ && \hspace{3cm} - \frac{1}{2} e_A{}^\P e_B{}^\O
(\p_\O e_\P{}_C  - \p_\P e_\O{}_C) e_\S{}^C \rlap{\ .} \eea

We perform a dimensional reduction, with the vielbein parametrized
by (\ref{vb}), by imposing the vanishing of derivatives $\p_\S$
for $\S \geq 3$. We also rescale $\psi$ by a power $f$ of the
determinant of the reduced part of the vielbein $M$: \be \hpsi =
e^{- \frac{1}{2} f \vec{s}.\vec{\phi}} \psi \rlap{\ .} \ee

After reassembling the various terms, we find that the reduced
Dirac Lagrangian can be written as \be
\begin{split}
\cL_D^\3 = \ & e^{(\frac{1}{2}+f)\vec{s}.\vec{\phi}} \he \bhpsi
\Sh{\hat{D}} \hpsi + \G( \frac{1}{2} f + \frac{1}{4} \D)
e^{(\frac{1}{2}+f)\vec{s}.\vec{\phi}}
\he \, \p_\mu(\vec{s}.\vec{\phi}) \, \bhpsi \gamma^\mu \hpsi \\
& + \frac{1}{8} e^{(\frac{1}{2}+f)\vec{s}.\vec{\phi}} \he \G(
M^{-1}{}_j{}^k \p_\mu M_k{}^i - M^{-1}{}_i{}^k \p_\mu M_k{}^j \D)
\bhpsi \gamma^\mu \gamma^{ij} \hpsi \\
& + \frac{1}{8} e^{f\vec{s}.\vec{\phi}} \he \, \he_m{}^\mu
e_n{}^\O \! \G( \p_\mu(\cA_\1{}_\O{}^j M^{-1}{}_j{}^k) -
\p_\O(\cA_\1{}_\mu{}^j M^{-1}{}_j{}^k) \D) \! M_k{}^i \, \bhpsi
\gamma^i \gamma^{mn} \hpsi
\end{split}
\ee where $\he$ is the determinant of the dreibein. Note that the
numerical matrices $\gamma^M$ (with $M$ an internal index) are
left unchanged in the reduction process, but this is not the case
for $\gamma^\S$ (with $\S$ a spacetime index). Indeed, $\g^{\S}$
for $\S = \m$ has to be understood as $e_M{}^{\mu} \g^M$ in $D$
dimensions, and as $\he_m{}^\mu \g^m $ in 3 dimensions.
Nevertheless, we do not put hats on three-dimensional
$\gamma$-matrices with a spatial index as no confusion should
arise.

Dualizing $\cA_\1$ according to (\ref{dual-a1}), we get \be
\begin{split}
\cL_D^\3 = \ & e^{(\frac{1}{2}+f)\vec{s}.\vec{\phi}} \he \bhpsi
\Sh{\hat{D}} \hpsi  + \G( \frac{1}{2} f + \frac{1}{4} \D)
e^{(\frac{1}{2}+f)\vec{s}.\vec{\phi}}
\he \, \p_\mu(\vec{s}.\vec{\phi}) \, \bhpsi \gamma^\mu \hpsi \\
& + \frac{1}{8} e^{(\frac{1}{2}+f)\vec{s}.\vec{\phi}} \he \G(
M^{-1}{}_j{}^k \p_\mu M_k{}^i - M^{-1}{}_i{}^k \p_\mu M_k{}^j \D)
\bhpsi \gamma^\mu \gamma^{ij} \hpsi \\
& + \frac{1}{8} e^{(1 + f) \vec{s}.\vec{\phi}} \he \,
\epsilon_{mnp} \he_p{}^\mu M^{-1}{}_i{}^j \p_\mu \chi_j \bhpsi
\gamma^i \gamma^{mn} \hpsi \rlap{\ .}
\end{split}
\ee

If we choose the scaling power of the spinor as \be f =
-\frac{1}{2} \rlap{\ ,} \ee the Lagrangian simplifies to \be
\begin{split}
\cL_D^\3 = \ & \he \bhpsi \Sh{\hat{D}} \hpsi  + \frac{1}{8} \he
\G( M^{-1}{}_j{}^k \p_\mu M_k{}^i - M^{-1}{}_i{}^k \p_\mu M_k{}^j
\D)
\bhpsi \gamma^\mu \gamma^{ij} \hpsi \\
& + \frac{1}{4} \he \, e^{\frac{1}{2} \vec{s}.\vec{\phi}}
M^{-1}{}_i{}^j \p_\mu \chi_j \bhpsi \gamma^\mu \hgamma \gamma^i
\hpsi
\end{split}
\label{cld} \ee where we have used the notation $\hat{\gamma} =
\gamma^0 \gamma^{D-2} \gamma^{D-1}$.

In fact, the three-dimensional Lagrangian can be rewritten using a
covariant derivative including a connection with respect to the
gauge group $SO(n+1)$: \be \cL_D^\3 = \he \bhpsi
\Sh{\nabla} \hpsi \ee with \be \nabla_\mu = \p_\mu - \frac{1}{4}
\hat{\omega}_{\mu,mn} \gamma^{mn} - \frac{1}{2} \cQ_{\mu,ij}
J^{ij} \label{3.18}\ee where $\cQ$ is the $SO(n+1)$
connection (\ref{Q-An}), acting on Dirac spinors through \be
J_{ij} = \frac{1}{2} \gamma^{ij}\, , \; \; \; \; \; \; \; \;
J_{i(n+1)} = \frac{1}{2} \hgamma \gamma^i \,  \; \; \; \; \; \;\;
\;  (i,j=1..n) \rlap{\ .} \ee

These matrices define a spinorial representation of
$SO(n+1)$. The commutations relations are indeed \be
\begin{array}{ccl}
\G[ \frac{1}{2} \gamma^{ij}, \frac{1}{2} \gamma^{kl} \D] &=& 0 \\
\G[ \frac{1}{2} \gamma^{ij}, \frac{1}{2} \gamma^{ik} \D] &=&
-\frac{1}{2}
\gamma^{jk} \\
\G[ \frac{1}{2} \hat{\gamma} \gamma^i, \frac{1}{2} \gamma^{jk} \D] &=& 0 \\
\G[ \frac{1}{2} \hat{\gamma} \gamma^i, \frac{1}{2} \gamma^{ij} \D]
&=&
\frac{1}{2} \hat{\gamma} \gamma^j \\
\G[ \frac{1}{2} \hgamma \gamma^i, \frac{1}{2} \hgamma \gamma^j \D]
&=& -\frac{1}{2} \gamma^{ij}
\end{array}
\ee where different indices are supposed to be distinct.
Equivalently, we can remark that $\hat{\gamma}$ commutes with
$\gamma^m$ for $m=0,D-2,D-1$ and anticommutes with $\gamma^i$ for
$i=1,...,n$. As we have also $\hat{\gamma}^2 = 1$, it follows that
$\gamma^a$'s for $a=1,...,n$ and $\hat{\gamma}$ generate an
internal $Spin(n+1)$ Clifford algebra, commuting with the
spacetime Clifford algebra generated by $\gamma^m, m=0,D-2,D-1$.

In other words, the $Spin(n+2,1)$ representation of Dirac fermions
in dimension $D=3+n$ is reduced to a $Spin(2,1) \times Spin(n+1)$
representation in dimension 3, ensuring that the Dirac fermions
are compatible with the hidden symmetry. Note that if $D$ is even,
one can impose chirality conditions on the spin $1/2$ field in $D$
dimensions. One gets in this way a chiral spinor of $SO(n+1)$
after dimensional reduction.

\subsection{Explicit Borel decomposition}
One may write the Lagrangian in the form (\ref{generalform}) by
making a full Borel parametrization of the matrix $M$. The algebra
element $\cG$ reads \be \cG = \frac{1}{2} \ud \vec{\phi} . \vec{H} +
\sum_{i<j} e^{ \frac{1}{2} \vec{b}_{ij}.\vec{\phi}} \cF_\1{}^i{}_j
e_{b_{ij}} + \sum_i e^{-\frac{1}{2} \vec{b}_i.\vec{\phi}} \cG_\1{}^i
e_{b_i} \ee where the dilaton vectors are given by \be \vec b_i = -
\vec s +2 \vec \g_i\ ,\qquad \vec b_{ij} = 2 \vec \g_i -2 \vec \g_j
\ee and where the ``field strengths" $\cF^i_{\1 j}$ and $\cG_\1{}^i$
(which are also the $SO(n+1)$ connections) are \be\qquad \cF^i_{\1
j} = \gamma^k{}_j\, d {\cal A}^i_{\0 k} , \; \; \; \; \; \cG_\1{}^i
= \gamma^j{}_i \ud \chi_j \label{fs0}\ee One has \be
e^{\vec{b}_i.\vec{\phi}} * \! \cF_\2{}^i = \cG_\1{}^i \ee with \be
\cF^i_{(2)} = \td\g^i{}_j d (\g^j{}_m {\cA}_\1^m ) \ee The positive
roots of $SL(n+1)$ are $(\vec b_{ij}, -\vec b_i)$ and the
corresponding root vectors $e_{b_{ij}}$ and $e_{b_i}$ are the
multiple commutators of the generators $e_i$ not involving $e_1$
(for $e_{b_{ij}}$) or involving $e_1$ (for $e_{b_i}$), i.e.,
$e_{b_{ii+1}} = e_{i+1}$ ($i = 1, \cdots, n-1$), $e_{b_{ij}} = [e_i,
[e_{i+1},[ ... [e_{j-2}, e_{j}] \cdots ] $ ($i, j = 2, \cdots, n$,
$i+1 < j$), $e_{b_i} = [e_1, e_{b_{2i}}]$ ($i\geq 3$), $e_{b_n} =
e_n$. These root vectors are such that the normalization factors
$N_\a$ are all equal to one.

The Lagrangian reads \bea {\cal L}_{D_n }^{(3)} &=& R\,
{*\oneone} - \ft12 {*d\vec\phi}\wedge d\vec\phi - \ft12 \sum_i
e^{-\vec b_i\cdot\vec\phi}\, {*\cG_{\1 i}}\wedge \cG_{\1 i} -\ft12
\sum_{i<j} e^{\vec b_{ij}\cdot\vec\phi}\, {*\cF^i_{\1 j}} \wedge
\cF^i_{\1 j}\no\\
&&+ \he \bhpsi \gamma^\mu \left( \p_\mu - \frac{1}{4}
\hat{\omega}_{\mu,mn} \gamma^{mn} - \ft14 \sum_{i<j} e^{\ft12 \vec
b_{ij}\cdot\vec\phi}\, {\cF^i_{\1 j}} \gamma^{ij} - \ft14 \sum_i
e^{-\ft12 \vec b_i \cdot\vec\phi}\, \cG_{\1 i} \hat{\gamma}
\gamma^{i}\right) \psi \hspace{.5cm}\label{anlag2} \eea

\section{$D_n$ case}
\label{dncase}
\subsection{Bosonic sector}

Following \cite{Cremmer:1999du}, we consider the gravitational
lagrangian ${\cal L}_E$ with an added three form field strength
$F_\3$ coupled to a dilaton field $\varphi$,
\be {\cal L} =R\, {*\oneone} - \ft12 {*d\varphi}\wedge d\varphi -
\ft12 e^{a\varphi}\, {*F_\3}\wedge F_\3 \label{dnlag} \ee
in the dimension $\Dm = n+2$, where the coupling constant $a$ is
given by $a^2=8/(\Dm-2)$. Upon toroidal reduction to $D=3$, this
yields the Lagrangian
\bea {\cal L}^{(3)} &=& R\, {*\oneone} - \ft12 {*d\vec\phi}\wedge
d\vec\phi - \ft12 \sum_i e^{\vec b_i\cdot\vec\phi}\,
{*\cF_\2^i}\wedge \cF_\2^i -\ft12 \sum_{i<j} e^{\vec
b_{ij}\cdot\vec\phi}\, {*\cF^i_{\1 j}} \wedge
\cF^i_{\1 j}\no\\
&&-\ft12 \sum_i e^{\vec a_i\cdot\vec\phi}\, {*F_{\2 i}}\wedge
F_{\2 i} -\ft12 \sum_{i<j} e^{\vec a_{ij}\cdot\vec\phi}\, {*F_{\1
ij}}\wedge F_{\1 ij}\ .\label{dnlag1} \eea
Note that here $\vec\phi$ denotes now the set of dilatons
$(\phi_1,\phi_2,\ldots, \phi_{{\sst \Dm}-3})$, augmented by
$\varphi$ (the dilaton in $\Dm$ dimensions) as a zeroth component;
$\vec\phi=(\varphi, \phi_1,\phi_2,\ldots, \phi_{{\sst \Dm}-3})$.
The dilaton vectors entering the exponentials in the Lagrangian
are given by \be \vec a_i = -2 \vec \g_i- \vec s\ ,\qquad \vec
a_{ij} = -2 \vec \g_i -2 \vec \g_j \ , \qquad  \vec b_i = - \vec s
+2 \vec \g_i\ ,\qquad \vec b_{ij} = 2 \vec \g_i -2 \vec \g_j \ . \
\label{dvec} \ee augmented by a zeroth component that is equal to
the constant $a$ in the case of $\vec a_i$ and $\vec a_{ij}$, and
is equal to zero in the case of $\vec b_i$ and $\vec b_{ij}$. The
field strengths are given by
\bea \cF^i_{(2)} &=& \td\g^i{}_j
d (\g^j{}_m {\cA}_\1^m )\ , \nn \\
\qquad \cF^i_{\1 j} &=&
\gamma^k{}_j\, d {\cal A}^i_{\0 k}\ , \nn \\
F_{{(2)}i} &=& \g^{k}{}_{i} (dA_{\sst{(1)}k} + \g^j{}_m
dA_{{(0)}kj} \wedge
\hA_\1^m), \nn\\
F_{\1 ij} &=& \g^k{}_i \g^m{}_j dA_{\0 km} \label{fs1}. \nn \eea
After dualising the 1-form potentials $\cA^i_\1$ and $A_{\1 i}$ to
axions $\chi_i$ and $\psi^i$ respectively, the three-dimensional
Lagrangian (\ref{dnlag1}) can written as the purely scalar
Lagrangian
\bea {\cal L}_{D_n }^{(3)} &=& R\, {*\oneone} - \ft12
{*d\vec\phi}\wedge d\vec\phi - \ft12 \sum_i e^{-\vec
b_i\cdot\vec\phi}\, {*\cG_{\1 i}}\wedge \cG_{\1 i} -\ft12
\sum_{i<j} e^{\vec b_{ij}\cdot\vec\phi}\, {*\cF^i_{\1 j}} \wedge
\cF^i_{\1 j}\no\\
&&-\ft12 \sum_i e^{-\vec a_i\cdot\vec\phi}\, {* G_\1^i}\wedge
G_\1^i -\ft12 \sum_{i<j} e^{\vec a_{ij}\cdot\vec\phi}\, {*F_{\1
ij}}\wedge F_{\1 ij}\ ,\label{dnlag2} \eea
where the dualised field strengths are given by
\bea e^{\vec{b}_i.\vec{\phi}} * \! \cF_\2{}^i \equiv \cG_\1{}^i
&=& \gamma^j{}_i
(d\chi_j - A_{\0 kj}\, d\psi^k)\ ,\no\\
G^i_\1 &=& e^{-\vec \g_i \cdot \vec \phi}M^i{}_j\, d\psi^j\ . \eea
Note the modification of $\cG_\1{}^i$ due to the coupling of
$\cA^i_\1$ to the $2$-form variables. The positive roots of $D_n$
are given by $(\vec b_{ij}, -\vec b_i, \vec a_{ij}, -\vec a_i)$ ,
the simple roots being $\vec a_{12}$, $\vec b_{i,i+1}$ ($i\le
n-1$) and $-\vec a_n$ \cite{Cremmer:1999du}. The three-dimensional
Lagrangian (\ref{dnlag2}) describes a $SO(n,n) / (SO(n)\times
SO(n) ) $ $\sigma$-model in the Borel gauge coupled to gravity.
The field strength of this $\sigma$-model is \be
\begin{split}
\cG = \frac{1}{2} \ud \vec{\phi} . \vec{H} + \sum_{i<j} e^{
\frac{1}{2} \vec{b}_{ij}.\vec{\phi}} \cF_\1{}^i{}_j e_{b_{ij}} +
\sum_i e^{-\frac{1}{2} \vec{b}_i.\vec{\phi}} \cG_\1{}^i e_{b_i}
\\
+ \sum_{i<j} e^{\frac{1}{2} \vec{a}_{ij}.\vec{\phi}} F_{\1 ij}
e_{a_{ij}} + \sum_{i} e^{-\frac{1}{2} \vec{a}_{i}.\vec{\phi}}
G_\1{}^{i} e_{a_i} \rlap{\ .}
\end{split}
\label{g-dn} \ee $\vec{h}$ is the vector of Cartan generators and
the notations $e_{a_i}, \ e_{a_{ij}}, e_{b_i}$ and $e_{b_{ij}}$
are explained in appendix \ref{appendixdn}.

\subsection{Fermions}

We want to add Dirac fermion in $D_{max}$, with a coupling which
reduces to $SO(2,1) \times (SO(n)\times SO(n))$. The coupling to
gravitational degrees of freedom is already fixed to the spin
connection by invariance under reparametrization; we know from the
first section that it reduces to the $SO(2,1)\times SO(n)$
connection in $D=3$. From the structure of the theory, we know
that the fermions must have linear couplings with the $3$-form
$F_3$. Indeed, the $D=3$ couplings must be of the following form
\be \he \bhpsi \gamma^{\mu} \G( \p_\mu - \frac{1}{4} \hat{\omega}
_{\mu,mn} \gamma^{mn} - \cQ_{\mu (\alpha)} J^{(\alpha)} \D) \psi
\ee where $\cQ$ can be read off from (\ref{g-dn}) above and the
$J^{(\a)}$'s are a representation of $SO(n) \times SO(n)$. The
possible Lorentz-covariant coupling of this kind are the Pauli
coupling and its dual, \be \ -\sqrt{-g} e^{\frac{1}{2} a
\varphi}\bar{\psi} { 1 \over 3!}(\a \g^{A B C} + \b \g^{ABC} \g)
F_{(3) \, ABC} \label{dnlagf0} \psi \ee where $\a$ and $\b$ are
arbitrary constants, which will be determined below. The dilaton
dependence is fixed so as to reproduce the roots $\vec{a}_i$ and
$\vec{a}_{ij}$ in the exponentials in front of the fermions in the
expressions below. The matrix $\g$ is the product of all gamma
matrices $\g = \g^0 \g^1 ... \g^{D-1}$. One has $\gamma^2 =
-(-1)^{[ {D \over 2} ]}$. Notice that in odd dimensions this
matrix is proportional to the identity and therefore we can put
$\b=0$ without loss of generality. Thus we add to the bosonic
lagrangian (\ref{dnlag}) the following term,
\be {\cal L}_{\psi} = \ \sqrt{-g} \bar{\psi} (
\g^{\m}\partial_{\m} - {1 \over 4} \o_{\m,mn} \g^{mn} - { 1 \over
3!}(\a \g^{A B C} + \b \g^{ABC} \g)  e^{\frac{1}{2} a
\varphi} F_{(3) \, ABC} \label{dnlagf} )
\psi \ee
Upon toroidal reduction to $D=3$, the last term of (\ref{dnlagf})
becomes, \be - { 1 \over 3!} \sqrt{-\hat{g}} \bar{\hpsi} \Big(
e^{\ft12 \vec a_i\cdot\vec\phi } 3 ( \a \g^{ab} \g^{i}+ \b \g^{ab
i} \g ) F_{(2)i \ ab} + e^{\ft12 \vec a_{ij}\cdot\vec\phi } 3!( \a
\g^a \g^{j} \g^{i} + \b \g^a \g^{j}\g^{i} \g ) F_{(1)ji \ a} \Big)
\hpsi \ee Let us dualize the 2 form field strengths.  By using the
relation $\e_{cab} \g^{ab}= 2 \g_c \hg $, we get for the
dimensional reduction of the whole lagrangian (\ref{dnlagf}) (with
dualisation of the $\cF^{(i)}$'s and using the results of the pure
gravitational case), \bea {\cal L}_{\psi}^{(3)} = \sqrt{-\hat{g}}
\bar{\hpsi} \g_c(\g^{\m}\partial_{\m} &-& {1 \over 4}
\hat{\o}_{\m,mn} \g^{mn} \nn \\&-& {1 \over 2} e^{-\ft12 \vec
a_i\cdot\vec\phi } \Gamma_{\vec a_i} G^c_i - {1 \over 2}e^{\ft12
\vec a_{ij}\cdot\vec\phi } \Gamma_{\vec a_{ij}} F^c_{(1) ij} \nn \\
&-& {1 \over 2}e^{-{1 \over 2}\vec b_i . \vec \phi} \Gamma_{b_i}
\cG^c_i - {1 \over 2}e^{{1 \over 2}\vec b_{ij} . \vec \phi }
\Gamma_{b_{ij}} \cF^c_{(1) ij}) \hpsi \label{lagdn3}\eea where
\bea \label{rep-dn}&& \Gamma_{\vec a_i} = 2(\a \hg \g^i + \b \hg
\g^i \g) , \; \; \; \; \;  \Gamma_{\vec a_{ij}} = 2(\a \g^i \g^j +
\b \g^i \g^j \g ),\nn \\ && \Gamma_{\vec b_i} = \ft12 \hg \g^i ,
\; \; \; \; \;  \; \; \; \; \Gamma_{\vec b_{ij}} = \ft12 \g^i
\g^j\eea

We have to compare this expression with \be \sqrt{-\hat{g}}
\bar{\hpsi} \g_c(\g^{\m}\partial_{\m} - {1 \over 4}
\hat{\o}_{\m,mn} \g^{mn} - \cQ ^{ c  ( \a)} J^{( \a)})\psi \ee
where $\cQ_{ \mu (\a)}$ are the coefficients of the $K(SO(n,n)) =
SO(n) \times SO(n)$ gauge field. From (\ref{g-dn}), we find that
\be
\begin{split}
\cQ = & \frac{1}{2} \sum_{i<j} e^{ \frac{1}{2}
\vec{b}_{ij}.\vec{\phi}} \cF_\1{}^i{}_j (e_{b_{ij}} + f_{b_{ij}})
+ \frac{1}{2} \sum_i e^{-\frac{1}{2} \vec{b}_i.\vec{\phi}}
\cG_\1{}^i (e_{b_i}+f_{b_i})
\\
&+ \frac{1}{2} \sum_{i<j} e^{\frac{1}{2} \vec{a}_{ij}.\vec{\phi}}
F_{\1 ijk} (e_{a_{ij}} + f_{a_{ij}}) + \frac{1}{2} \sum_{i}
e^{-\frac{1}{2} \vec{a}_{i}.\vec{\phi}} G_\1{}^{i} (e_{a_i} +
f_{a_i}) \rlap{\ .}
\end{split}
\ee The commutation relations of the $k_{(\a)}$'s are explicitly
given in appendix \ref{appendixdn}. One has to fix the values of
$\a$ and $\b$ such that the generators $\Gamma_{\vec a_i}$,
$\Gamma_{\vec b_i}, \ \Gamma_{\vec a_{ij}}$ and $\Gamma_{\vec
b_{ij}} $ obey these same commutation relations. The conditions we
found are  \be \label{cond-dn} \a^2 + \b^2 \g^2 = {1 \over 16} \,
, \; \; \; \; \; \;  \;  \a \b = 0. \ee

In odd dimension, we have set $\b=0$. This implies $\a = \pm
\frac{1}{4}$. We get for each choice of $\a$ a representation
which is trivial for either the left or the right $SO(n)$
factor of the compact gauge group. With $\b=0$, (\ref{rep-dn})
generates indeed $SO(n)$, as our analysis of the
gravitational sector has already indicated.

In even dimension, the choices $ \b = 0, \a = \pm \frac{1}{4}$ are
still solutions to (\ref{cond-dn}), but in addition one can have
$\a =0$, $\b = \pm \frac{\iota}{4} $, where the constant $\iota$
is $1$ or $i$ such that $(\iota \g)^2 = 1$. In this case,
(\ref{rep-dn}) combines with the gravitational $SO(n)$ to
give a $SO(n) \times SO(n)$ representation which
is nontrivial on both factors. The two factors
$SO(n)_\pm$ are generated in the spinorial space by the
matrices \be
\begin{array}{c}
\frac{1}{4} (1 \pm \iota \gamma) \gamma^{ij} \\
\frac{1}{4} (1 \pm \iota \gamma) \hgamma \gamma^i
\end{array}
\ee The (reduced) gravitational sector is given by the diagonal
$SO(n)$. If one imposes a chirality condition in $\Dm$
dimensions, the solution with $\b = 0$ and the solution with $\a =
0$ are of course equivalent and the representation is
trivial on one of the $SO(n)$.

This completes the proof that the Dirac spinors are compatible
with the $D_n$ hidden symmetry.

\section{$E_n$ sequence}
\label{e8case}

\subsection{$E_8$ -- bosonic}

We consider now the bosonic part of 11-dimensional supergravity,
\ie gravity coupled to a 3-form in 11 dimensions with the specific
value of the Chern-Simons term dictated by supersymmetry.  We
denote the 3-form $A_\3$ and its field strength $F_\4 = \ud A_\3$.

The Lagrangian is \cite{Cremmer:1978km} \be \cL = R\, {*\oneone} -
\frac{1}{2}  *\! F_\4 \w F_\4 - \frac{1}{3!} F_\4 \w F_\4 \w A_\3
\rlap{\ .} \label{e8-bos} \ee Prior to dualization, the 3-form
term of the Lagrangian reduces in three dimensions to \bea
\hat{\cL}_3 &=& -\frac{1}{2} \sum_{i<j<k}
e^{\vec{a}_{ijk}.\vec{\phi}} (F_\1{}_{ijk})^2 - \frac{1}{4} \he
\sum_{i<j} e^{\vec{a}_{ij}.\vec{\phi}} (F_\2{}_{ij})^2 \nonumber
\\ && \hspace{2cm} - \frac{1}{144} \ud A_\0{}_{ijk} \w \ud A_\0{}_{lmn} \w
A_\1{}_{pq} \epsilon^{ijklmnpq} \label{sugra-3} \eea In addition
to the gravitational degrees of freedom described in section
\ref{gravity}, we have 56 scalars $A_\0{}_{ijk}$ and 28 1-forms
$A_\1{}_{ij} = A_{\mu (i+2)(j+2)} \ud x^\mu$, with $i,j,k = 1..8$.
The reduced field strength are defined as \bea
F_\1{}_{ijk} &=& \gamma^l{}_i \gamma^m{}_j \gamma^n{}_k \, \ud A_\0{}_{lmn} \\
F_\2{}_{ij} &=& \gamma^k{}_i \gamma^l{}_j \G( \ud A_\1{}_{kl} -
\gamma^m{}_n \, \ud A_\0{}_{klm} \w \cA_\1{}^n \D) \label{F2}
\rlap{\ .} \eea The 1-forms $A_\1{}_{ij}$ are then dualized into
scalars $\lambda^{kl}$: \be e^{\vec{a_{ij}}.\vec{\phi}} * \!
F_\2{}_{ij} = G_\1{}^{ij} = (\gamma^{-1})^i{}_k
(\gamma^{-1})^j{}_l \G( \ud \lambda^{kl} +\frac{1}{72} \ud A_{\0
mnp} A_{\0 qrs} \epsilon^{klmnpqrs} \D) \rlap{\ .} \label{dual-f2}
\ee Moreover, the gravitational duality relation (\ref{dual-a1})
has to be modified to take into account the 3-form degrees of
freedom \be e^{\vec{b}_i.\vec{\phi}} * \! \cF_\2{}^i = \cG_\1{}^i
= \gamma^j{}_i \G( \ud \chi_j - \frac{1}{2} A_{\0 jkl} \ud
\lambda^{kl} - \frac{1}{432} \ud A_{\0 klm} A_{\0 npq} A_{\0 rsj}
\epsilon^{klmnpqrs} \D) \rlap{\ .} \ee

Taking all this into account, the full 3-dimensional Lagrangian
can be written as \be
\begin{split}
\hat{\cL} = R\, {*\oneone} - \frac{1}{2} \ud \vec{\phi} \dot{\w}
\!*\!\ud \vec{\phi} - \frac{1}{2} \he \sum_{i<j}
e^{\vec{b}_{ij}.\vec{\phi}} \cF_\1{}^i{}_j \w \!* \cF_\1{}^i{}_j
-\frac{1}{2} \he \sum_i e^{-\vec{b}_i.\vec{\phi}} \cG_\1{}^i \w
\!* \cG_\1{}^i
\\
-\frac{1}{2} \he \sum_{i<j<k} e^{\vec{a}_{ijk}.\vec{\phi}} F_{\1
ijk} \w \!*\! F_{\1 ijk} -\frac{1}{2} \he \sum_{i<j}
e^{-\vec{a}_{ij}.\vec{\phi}} G_\1{}^{ij} \w \!* G_\1{}^{ij}
\end{split}
\ee which describes a $E_{8(8)}/SO(16)$ $\sigma$-model coupled to
gravity \cite{CJ0,Marcus:1983hb,Cremmer:1997ct}, in the Borel
gauge, with field strength \be
\begin{split}
\cG = \frac{1}{2} \ud \vec{\phi} . \vec{H} + \sum_{i<j} e^{
\frac{1}{2} \vec{b}_{ij}.\vec{\phi}} \cF_\1{}^i{}_j e_{ij} + \sum_i
e^{-\frac{1}{2} \vec{b}_i.\vec{\phi}} \cG_\1{}^i e_i
\\
+ \sum_{i<j<k} e^{\frac{1}{2} \vec{a}_{ijk}.\vec{\phi}} F_{\1 ijk}
\tilde{e}_{ijk} + \sum_{i<j} e^{-\frac{1}{2}
\vec{a}_{ij}.\vec{\phi}} G_\1{}^{ij} \tilde{e}_{ij} \rlap{\ .}
\end{split}
\label{g-e8} \ee

The explicit expressions for the couplings $\vec{a}_{ijk}$ and
$\vec{a}_{ij}$ are (see \cite{Cremmer:1997ct}) \be \vec{a}_{ijk} =
-2(\vec{\gamma}_i + \vec{\gamma}_j + \vec{\gamma}_k), \; \; \;
\vec{a}_{ij} = -2 (\vec{\gamma}_i + \vec{\gamma}_j) -\vec{s}. \ee
The positive roots are $\vec{b}_{ij}$, $- \vec{b}_{i}$,
$\vec{a}_{ijk}$ and $-\vec{a}_{ij}$.  The elements $e_{ij}$
($i<j$), $e_i$, $\te_{ijk}$ and $\te_{ij}$ (with antisymmetry over
the indices for the two last cases) are the raising operators.
Note that $e_{ij}$ and $e_i$ generate the $SL(9)$ subalgebra
coming from the gravitational sector. In addition, there are
lowering operators $f_{ij}$, $f_i$, $\tf_{ijk}$ and $\tf_{ij}$. We
give all the commutation relations in that basis of $E_8$ in
appendix \ref{E8}.

\subsection{$E_8$ -- fermions}
\label{e8}

The maximal compact subgroup of $E_8$ is $SO(16)$;
its generators are given in appendix \ref{E8}. We want to add Dirac
fermions in $D=11$, with a coupling which reduces to a
$SO(2,1) \times SO(16)$-covariant derivative in
three dimensions. The coupling to gravitational degrees of freedom
is already fixed to the spin connection by invariance under
reparametrizations; we know from the first section that it reduces
to the relevant $SO(2,1) \times SO(9)$ connection
in $D=3$.

{}From the structure of the reduced theory, we know that the
fermion must have a linear coupling to the 4-form $F_\4$. The only
Lorentz-covariant coupling of this kind for a single Dirac fermion
in $D=11$ is a Pauli coupling \be e a \frac{1}{4!} \bpsi F_{\mu
\nu \rho \sigma} \gamma^{\mu \nu \rho \sigma} \psi \ee where $a$
is a constant. Indeed in odd dimensions, the product of all
$\gamma$ matrices is proportional to the identity, so the dual
coupling is not different: \be \frac{1}{7!} (*F)_{\mu_1 \ldots
\mu_7} \gamma^{\mu_1 \ldots \mu_7} = \frac{1}{4!} F_{\mu_1 \ldots
\mu_4} \gamma^{\mu_1 \ldots \mu_4} \rlap{\ .} \ee Thus we add to
the bosonic Lagrangian (\ref{e8-bos}) the fermionic term \be
\cL_\psi = e \bpsi \G(\gamma^\mu \p_\mu - \frac{1}{4}
\omega_\mu{}^{ab} \gamma^\mu \gamma^{ab} - \frac{1}{4!} a
F_{\mu\nu\rho\sigma} \gamma^{\mu\nu\rho\sigma} \D) \psi \ee where
$\gamma$ matrices with greek, curved indices must be understood as
$\gamma^\mu = e_a{}^\mu \gamma^a$.

Dimensional reduction to $D=3$ leads to \be \cL_\psi^\3 = \cL_D^\3
- \he a \frac{1}{3!} e^{\frac{1}{2} \vec{a}_{ijk}.\vec{\phi}}
F_{\1 \mu ijk} \bhpsi \gamma^\mu \gamma^{ijk} \hpsi - \he a
\frac{1}{2.2} e^{\frac{1}{2} \vec{a}_{ij}.\vec{\phi}} F_{\2 \mu\nu
ij} \bhpsi \gamma^{\mu\nu} \gamma^{ij} \hpsi \ee where $\cL_D^\3$
is part not containing the 3-form computed previously in
(\ref{cld}), and with the same rescaling of $\psi$ into $\hpsi$.
Dualisation (\ref{dual-f2}) of $F_{\2 ij}$ can be written as \be
\frac{1}{2} e^{\frac{1}{2} \vec{a}_{ij}.\vec{\phi}} F_{\2 \mu\nu
ij} \gamma^{\mu\nu} = e^{-\frac{1}{2} \vec{a}_{ij}.\vec{\phi}}
G_{\1 \mu ij} \gamma^\mu \hgamma \rlap{\ .} \ee It gives the fully
dualised fermionic term \be \cL_\psi^\3 = \cL_D^\3 - \he a
\frac{1}{3!} e^{\frac{1}{2} \vec{a}_{ijk}.\vec{\phi}} F_{\1 \mu
ijk} \bhpsi \gamma^\mu \gamma^{ijk} \hpsi - \he a \frac{1}{2}
e^{-\frac{1}{2} \vec{a}_{ij}.\vec{\phi}} G_{\1 \mu ij} \bhpsi
\gamma^\mu \hgamma \gamma^{ij} \hpsi \rlap{\ .} \label{l3-e8} \ee

We have to compare this expression to \be \he \bhpsi \gamma^\mu
\G( \p_\mu - \frac{1}{4} \hat{\omega} _{\mu,mn} \gamma^{mn} -
\cQ^{\mu (\alpha)} J^{(\alpha)} \D) \psi . \ee {}From
(\ref{g-e8}), we have \be
\begin{split}
\cQ = & \frac{1}{2} \sum_{i<j} e^{ \frac{1}{2}
\vec{b}_{ij}.\vec{\phi}} \cF_\1{}^i{}_j (e_{ij} + f_{ij}) +
\frac{1}{2} \sum_i e^{-\frac{1}{2} \vec{b}_i.\vec{\phi}}
\cG_\1{}^i (e_i+f_i)
\\
&+ \frac{1}{2} \sum_{i<j<k} e^{\frac{1}{2}
\vec{a}_{ijk}.\vec{\phi}} F_{\1 ijk} (\te_{ijk} + \tf_{ijk}) +
\frac{1}{2} \sum_{i<j} e^{-\frac{1}{2} \vec{a}_{ij}.\vec{\phi}}
G_\1{}^{ij} (\te_{ij} + \tf_{ij}) \rlap{\ .}
\end{split}
\ee In fact, we have precisely the correct gauge connection that
appears in (\ref{l3-e8}). We have only to check that the products
of gamma matrices that multiply the connection in (\ref{l3-e8})
satisfy the correct commutation relations. Using the commutation
relations of the compact generators  $k_{ij} = e_{ij} + f_{ij}$ $
(i<j)$, $k_i = e_i+f_i$, $\tk_{ijk} = \te_{ijk} + \tf_{ijk}$,
$\tk_{ij} = \te_{ij} + \tf_{ij}$ given in appendix \ref{E8} we
find that the coupling constant must be $a=-\frac{1}{2}$. The
spinorial generators are then given by \be
\begin{array}{ll}
k_{ij}: & \frac{1}{2} \gamma^{ij} \\
k_i: & \frac{1}{2} \hgamma \gamma^i \\
\tk_{ijk}: & -\frac{1}{2} \gamma^{ijk} \\
\tk_{ij}: & -\frac{1}{2} \hgamma \gamma^{ij}
\end{array}
\ee (we define $k_{ij} = - k_{ji} = -e_{ji} - f_{ji} $ for  $
i>j$). We have recovered the well known feature that the spinorial
representation of $\mathrm{so}(9)$ is the vector representation of
$\mathrm{so}(16)$ (see \cite{Keur3} for more on this).

We see also that $E_{8(8)}$-invariance forces one to
introduce the covariant Dirac operator
\be \gamma^\mu D_\mu \psi =
\gamma^\mu (\p_\mu - \frac{1}{4} \omega_\mu{}^{ab} \gamma^{ab} )
\psi + \frac{1}{2 . 4!} F_{\mu\nu\rho\sigma} \gamma^{\mu
\nu\rho\sigma} \psi \ee
for the Dirac field. This is exactly the
same which appears in $D=11$ supergravity, but it is
obtained in that context from supersymmetry.

\subsection{IIB}

The oxidation of the $E_{8(8)}/SO(16)$ coset
theory has another endpoint, in $D=10$: the bosonic sector of type
\emph{IIB} supergravity. There is no manifestly covariant
Lagrangian attached to this theory.  Indeed, the theory contains a
selfdual 4-form, which has no simple (quadratic) manifestly
covariant Lagrangian (although it does admit a quadratic non
manifestly covariant Lagrangian \cite{HHTT}, or a non polynomial
manifestly covariant Lagrangian \cite{Pasti:1996vs}). In spite of
the absence of a  covariant Lagrangian, the equations of motion
are covariant and one may address the following question: is there
a ``covariant Dirac operator'' for fermions in $D=10$ which reduces to
the same $SO(16)$ covariant derivative in $D=3$?

Following the notations of \cite{Cremmer:1998px}, we have for this
theory,  in addition to the metric, a dilaton $\phi$, an other
scalar $\chi$, two 2-forms $A_\2^1$ and $A_\2^2$ with field strength
$F_\3^1$ and $F_\3^2$, and a 4-form $B_\4$ with selfdual field
strength $H_\5$.

If it exists, the $D=10$ ``covariant Dirac operator'' would have the
form \be
\begin{split}
\gamma^\mu \nabla_\mu =  \gamma^\mu \p_\mu - \frac{1}{4} \omega_\mu{}^{ab}
\gamma^\mu \gamma^{ab} -
e^\phi \p_\mu \chi (a + \tilde{a} \gamma) \gamma^\mu - \frac{1}{3!}
e^{\frac{1}{2}\phi} F^1_{\mu \nu\rho}
(b + \tilde{b} \gamma) \gamma^{\mu\nu\rho} \\
- \frac{1}{3!} e^{-\frac{1}{2}\phi} F^2_{\mu \nu\rho} (c + \tilde{c}
\gamma) \gamma^{\mu\nu\rho} - \frac{1}{5!} H_{\mu\nu\rho\sigma\tau} f
\gamma^{\mu\nu\rho\sigma\tau} \rlap{\ .} \label{cd-iib}
\end{split}
\ee $\gamma = \gamma^{11}$ is the product of the ten $\gamma^i$
matrices. As $H_\5$ is selfdual, the dual term \be
H_{\mu\nu\rho\sigma\tau} \gamma\gamma^{\mu\nu\rho\sigma\tau} =
(*H)_{\mu\nu\rho\sigma\tau} \gamma^{\mu\nu\rho\sigma\tau} \ee is
already taken into account. The powers of the dilaton are fixed so
that the field strength give the expected fields in $D=3$.

Now, the axion term $e^\phi \partial_\mu \chi$ is the connection
for the $SO(2)$-subgroup of the $SL(2)$ symmetry
present in 10 dimensions.  Under $SO(2)$-duality,  the
two two-forms rotate into each other. So, the commutator of the
generator $(a + \tilde{a} \gamma)$ multiplying the connection
$e^\phi \partial_\mu \chi$ with the generators $(b + \tilde{b}
\gamma) \gamma^{\nu\rho}$ multiplying the connection
$e^{\frac{1}{2}\phi} F^1_{\mu \nu\rho}$ should reproduce the
generator $(c + \tilde{c} \gamma) \gamma^{\nu\rho}$ multiplying
the connection $e^{\frac{1}{2}\phi} F^2_{\mu \nu\rho}$. But one
has $[(a + \tilde{a} \gamma), (b + \tilde{b} \gamma)
\gamma^{\nu\rho}] = 0$, leading to a contradiction.

The problem just described comes from the fact that we have taken a
single Dirac fermion.  Had we taken instead two Weyl fermions, as it
is actually the case for type IIB supergravity, and assumed that
they transformed appropriately into each other under the
$SO(2)$-subgroup of the $SL(2)$ symmetry, we could have constructed
an appropriate covariant derivative.  This covariant derivative is
in fact given in \cite{westschwarz}, to which we refer the reader.
The $SO(2)$ transformations rules of the spinors ---~as well as the
fact that they must have same chirality in order to transform indeed
non trivially into each other ---~follow from $E_8$-covariance in 3
dimensions.

\subsection{$E_7$ case}

The $E_{7(7)}$ exceptional group is a subgroup of $E_{8(8)}$. As a
consequence, the $D=3$ coset $E_{7(7)}/SU(8)$ can be seen as a
truncation of the $E_{8(8)}/SO(16)$ coset theory. In fact, this
truncation can be made in higher dimension \cite{Cremmer:1999du}.
One can truncate the $D=9$ reduction of the gravity + 3-form
theory considered in the last section. If one does not worry about
Lagrangian, one can go one dimension higher and view the theory as
the truncation of the bosonic sector of type \emph{IIB}
supergravity in which one keeps only the vielbein and the chiral
4-form.

In $D=9$, the coupling to fermions obtained in section \ref{e8} is
truncated in a natural way: the components of the covariant
Dirac operator acting on fermions are the various fields of the
theory, so some of them just disappear with the truncation. The
symmetry of the reduced $D=3$ theory is thus preserved: the
fermions are coupled to the bosonic fields through a
$SU(8)$ covariant derivative, the truncation of the
$SO(16)$ covariant derivative of the $E_8$ case.

The question is about oxidation to $D=10$. Can we obtain this
truncated covariant Dirac operator from a covariant Dirac operator of the
$D=10$ theory? For the reasons already exposed, if it exists, this operator
would act on Dirac fermions as \be \gamma^\mu
\nabla_\mu = \gamma^\mu \p_\mu - \frac{1}{4} \omega_\mu{}^{ab}
\gamma^\mu \gamma^{ab} - a \frac{1}{5!} H_{\mu\nu\rho\sigma\tau}
\gamma^{\mu\nu\rho\sigma\tau} \ee where we have denoted by $H$ the
selfdual field strength.

With notations analogous to the $E_8$ case, we can write
the $D=3$ reduction of the covariant Dirac operator as \be
\begin{split}
\gamma^\mu \nabla_\mu = \gamma^\mu \p_\mu - \frac{1}{4}
\tilde{\omega}_\mu{}^{ab} \gamma^\mu \gamma^{ab} &- \frac{1}{4}
e^{\frac{1}{2}\vec{b}_{i}.\vec{\phi}} \cF_{\2 \mu\nu i}
\gamma^{\mu\nu} \gamma^i - \frac{1}{4}
e^{\frac{1}{2}\vec{b}_{ij}.\vec{\phi}}
\cF_{\1 \mu ij} \gamma^\mu \gamma^{ij} \\
&-a  \frac{1}{2.3!} e^{\frac{1}{2}\vec{a}_{ijk}.\vec{\phi}} H_{\2
\mu\nu ijk} \gamma^{\mu\nu} \gamma^{ijk} - a \frac{1}{4!}
e^{\frac{1}{2}\vec{a}_{ijkl}.\vec{\phi}} H_{\1 \mu ijkl}
\gamma^{\mu} \gamma^{ijkl} \rlap{\ .}
\end{split}
\ee Because of the selfduality of $H$, the 2-forms $H_\2$ and the
1-forms $H_\1$ are in fact dual. Using $\gamma = \gamma^0 \gamma^1
\ldots \gamma^9$, the covariant Dirac operator turns into \be
\begin{split}
\gamma^\mu \nabla_\mu = \gamma^\mu \p_\mu + \frac{1}{4}
\tilde{\omega}_\mu{}^{ab} \gamma^\mu \gamma^{ab} &+ \frac{1}{2}
e^{-\frac{1}{2}\vec{b}_{i}.\vec{\phi}} \cG_{\1 \mu i} \gamma^{\mu}
\hgamma \gamma^i + \frac{1}{2.2}
e^{\frac{1}{2}\vec{b}_{ij}.\vec{\phi}}
\cF_{\1 \mu ij} \gamma^\mu \gamma^{ij} \\
&+ a \frac{1}{4!} e^{\frac{1}{2}\vec{a}_{ijkl}.\vec{\phi}} H_{\1
\mu ijkl} \gamma^{\mu} (1+\gamma) \gamma^{ijkl} \rlap{\ .}
\end{split}
\label{cd-su8} \ee

The embedding of $E_{7(7)}/SU(8)$ in
$E_{8(8)}/SO(16)$ gives the following
identifications: \bea
H_{\1 1ijk} &=& F_{\1 (i+1)(j+1)(k+1)} \no\\
H_{\1 ijkl} &=& -\frac{1}{2} \epsilon^{12(i+1)(j+1)(k+1)(l+1)mn}
G_{\1 mn}
\no\\
\cF_{\1 1i} &=& F_{\1 12(i+1)} \no\\
\cG_{\1 1} &=& - G_{\2 12} \no\\
\cF_{\1 ij} &=& \cF_{\1 (i+1)(j+1)} \no\\
\cG_{\1 i} &=& \cG_{\1 (i+1)} \eea with $2 \leq i,j,k,l \leq 7$.
We thus have to check that the matrices in (\ref{cd-su8}) form the
following representation: \be
\begin{array}{rcl}
a (1+\gamma) \gamma^{1ijk} &\sim& \tk_{(i+1)(j+1)(k+1)} \\
a (1+\gamma) \gamma^{ijkl} &\sim&\frac{1}{2}
\epsilon^{12(i+1)(j+1)(k+1)(l+1)mn} \tk_{mn}\\
\frac{1}{2}\gamma^{1i} &\sim& \tk_{12(i+1)} \\
\frac{1}{2}\hgamma\gamma^1 &\sim& -\tk_{12} \\
\frac{1}{2}\gamma^{ij} &\sim& k_{(i+1)(j+1)} \\
\frac{1}{2}\hgamma\gamma^i &\sim& k_{(i+1)} \rlap{\ .}
\end{array}
\label{alg-su8} \ee This is true if and only if \be -4 a^2
(1+\gamma) = \frac{1}{2} \rlap{\ .} \ee This has to be understood
as an identity between operators acting on fermions. In fact, this
means that we must restrict to Weyl spinors, with $\gamma= +1$
when acting on them. Due to the even number of $\gamma$ matrices
involved in all generators in (\ref{alg-su8}), the
$\mathrm{su}(8)$ algebra preserves the chirality of spinors. We
get in addition the value of the coupling constant: \be a = \pm
\frac{i}{4} \rlap{\ .} \ee

\subsection{$E_6$  case}

The $E_6$ case is more simple. One has a Lagrangian in all
dimensions. In dimension 3, the scalar coset is $E_{6(6)}/Sp(4)$.
Maximal oxidation is a $D=8$ theory with a 3-form, a
dilaton and an axion (scalar) \cite{Cremmer:1999du}. It can be
seen as a truncation of the $E_8$ case in all dimensions. In the
compact subalgebra of $\mathrm{so}(16)$ given in (\ref{so16}), one
should remove generators with one or two indices in $ \{1,2,3\} $
while keeping $\tk_{123}$.

In fact, all the matrices involved in the Dirac representation can
be expressed in terms of a $D=8$ Clifford algebra. For most
generators, it is trivial to check that they involve only
$\gamma^i$ matrices with $i\neq 1,2,3$. The single nontrivial case
is $\tk_{123}$ which is represented by $-\frac{1}{2} \gamma^{123}$
in the eleven-dimensional Clifford algebra. But from the fact that
$\gamma^{(10)} = \gamma^0 \gamma^1 \ldots \gamma^9$, we can write
$\gamma^{123}$ as the product of all other $\gamma$ matrices:
$\gamma^{123} = \gamma^{0456789(10)}$. As a consequence, the $D=8$
Clifford algebra is sufficient to couple a Dirac fermion to this
model: we can couple a single $D=8$ Dirac fermion.

\section{$G_2$ case}
\label{g2case}

\subsection{Bosonic sector}

Let us consider the Einstein-Maxwell system in $D=5$, with the
$FFA$ term prescribed by supersymmetry \cite{ChamNic},
\be {\cal L}_5 = R\, {*\oneone} - \ft12 {*F_\2}\wedge F_\2
+\ft1{3\sqrt3} F_\2\wedge F_\2\wedge A_\1\ .\label{g2lag} \ee
This action is known to be relevant to $G_2$
\cite{Miz1,Cremmer:1999du}. Upon reduction to $D=3$, the
Lagrangian is \cite{Cremmer:1999du}
\bea {\cal L} &=& R\, {*\oneone} -\ft12 {*d\vec\phi}\wedge
d\vec\phi - \ft12 e^{\phi_2 -\sqrt3\phi_1}\, {*\cF^1_{\1 2}}\wedge
\cF^1_{\1 2}
-\ft12 e^{\fft2{\sqrt3}\phi_1}\, {*F_{\1 1}}\wedge F_{\1 1} \nn\\
&&-\ft12 e^{\phi_2 -\ft1{\sqrt3}\phi_1}\, {*F_{\1 2}}\wedge F_{\1
2} - \ft12 e^{-\phi_2 -\sqrt3\phi_1}\, {*\cF_\2^1}\wedge \cF_\2^1
\label{d5einstmax}\\
&&- \ft12 e^{-2\phi_2}\, {*\cF_\2^2}\wedge \cF_\2^2 -\ft12
e^{-\phi_2 -\ft1{\sqrt3}\phi_1}\, {*F_\2}\wedge F_\2 +
\ft{2}{\sqrt3} dA_{\0 1}\wedge dA_{\0 2}\wedge A_\1\ .\nn \eea
After dualising the vector potentials to give axions, there will
be six axions, together with the two dilatons. The dilaton vectors
$\vec\a_1=(-\sqrt3,1)$ and $\vec\a_2=(\ft2{\sqrt3}, 0)$,
corresponding to the axions $\cA^1_{\0 2}$ and $A_{\0 1}$, are the
simple roots of $G_2$, with the remaining dilaton vectors
expressed in terms of these as
\be (-\ft1{\sqrt3}, 1)=\vec\a_1+\vec \a_2\ ,\quad
(\ft1{\sqrt3},1)= \vec\a_1 + 2\vec\a_2\ ,\quad (\sqrt3,1)=
\vec\a_1+3\vec\a_2\ ,\quad (0,2)=2\vec\a_1+3\vec a_3\ . \ee
The resulting $D=3$ lagrangian is a $G_2 / SO(4)$ $\sigma$-model
coupled to gravity. The field strength of this $\sigma$-model is
\be
\begin{split}
\cG = \frac{1}{2} \ud \vec{\phi} . \vec{H} + e^{ \frac{1}{2}
\vec{\a}_{1}.\vec{\phi}} \cF_\1{}^1{}_2 \e_{1} + e^{\frac{1}{2}
(\vec \a_1 + 3 \vec \a_2).\vec{\phi}} \cG_\1{}^1 \e_{5}+
e^{\frac{1}{2} (2 \vec \a_1 + 3 \vec \a_2).\vec{\phi}} \cG_\1{}^2
\e_{6}
\\
+ e^{\frac{1}{2} \vec{\a}_{2}.\vec{\phi}} F_{\1 1} \e_{2}+
e^{\frac{1}{2} (\vec{\a}_{1} + \vec \a_2).\vec{\phi}} F_{\1 2}
\e_{3} + e^{-\frac{1}{2} (\vec{\a}_1 + 2 \vec \a_2).\vec{\phi}}
G_\1{} \e_{4} \rlap{\ .}
\end{split}
\label{g-dn2} \ee where $G_\1 $ is the dual of $F_2$, and the
notation $\e_1$,   $\e_2$, $\e_3$,  $\e_4$,  $\e_5 $ and $\e_6$ is
explained in the appendix.

\subsection{Fermions}

We want to add Dirac fermion in $D=5$, with a coupling which in
the $D=3$ reduction is covariant with respect to $SO(1,2) \times
SO(4)$. From what we have already learned, this
should be possible,  with a representation which is trivial on one
of the two $SU(2)$ factors of $SO(4) \simeq (SU(2) \times SU(2))/\ZZ_2$,
since we have already seen in the
analysis of the gravitational sector that the Clifford algebra
contains $SO(1,2) \times SU(2)$ representations.

To check if we can indeed derive such a representation from a
consistent $D=5$ coupling, we add to the lagrangian \ref{g2lag} a
Dirac fermion with a Pauli coupling,
\be {\cal L}_{\psi} = \ \sqrt{-g} \bar{\psi} (
\g^{\m}\partial_{\m} - {1 \over 4} \o_{\m,mn} \g^{mn} - { 1 \over
2}\a \g^{\m \n } F_{(2) \ \m \n} \label{g2lagf} ) \psi \ee
where $\a$ is a coupling constant which will be determined below.
Upon toroidal reduction to $D=3$, the last term of (\ref{g2lagf})
becomes, \bea { \a \over 2} \sqrt{-\hat{g}} \bar{\hpsi} (e^{-\ft12
(\vec \a_1 + 2 \vec \a_2)\cdot\vec\phi } \g^{ab} F_{(2) \ ab} + 2
e^{\ft12 \vec \a_2 \cdot\vec\phi } \g^a \g^{1} F_{(1)1 \ a} + 2
e^{\ft12 (\vec \a_1 + \vec \a_2 )\cdot\vec\phi } \g^a \g^{2}
F_{(1)2 \ a}) \hpsi \eea Let us dualize the 2 form field
strengths.  We get for the dimensional reduction of the whole
lagrangian (\ref{g2lagf}), \bea {\cal L}_{\psi}^{(3)} &=&
\sqrt{-\hat{g}} \bar{\hpsi} \g_c  ( \g^{\m}\partial_{\m} - {1
\over 4} \hat{\o}_{\m,mn} \g^{mn} \nn
\\&& \hspace{.5cm} -{1 \over 2} e^{\ft12 (\vec \a_1 + 2 \vec \a_2)\cdot\vec\phi
} \Gamma_{4} G^c - {1 \over 2}e^{\ft12 \vec \a_2\cdot\vec\phi }
\Gamma_{2} F^c_{(1) 1}
- {1 \over 2}e^{\ft12 (\vec \a_1 + \vec \a_2)\cdot\vec\phi }
\Gamma_{3} F^c_{(1) 2} \nn \\
&& \hspace{.5cm} - {1 \over 2}e^{{1 \over 2}(\vec \a_1 + 3 \vec
\a_2). \vec \phi} \Gamma_{5} \cG^c_1 -{1 \over 2}e^{{1 \over 2}(2
\vec \a_1 + 3 \vec \a_2). \vec \phi} \Gamma_{6} \cG^c_2 - {1 \over
2}e^{{1 \over 2}\vec \a_1 . \vec \phi } \Gamma_{1} \cF^{c \
1}_{(1) \ 2}) \hpsi \eea where $\Gamma_1 = {1\over 2} \g^{12}, \
\Gamma_2 = 2 \a \g^1, \ \Gamma_3 = 2 \a \g^2, \ \Gamma_4 = 2 \a
\hat{\g}, \ \Gamma_5 = {1\over 2 } \hat{\g} \g^1 $ and $\Gamma_6 =
{1\over 2} \hat{\g} \g^2$. Notice that $\hat{\g} = -i \g^{12}$
because the product of all gamma matrices $\g^0\g^1\g^2\g^3 \g^4 =
\hat{\g} \g^1\g^2$ in $D=5$ can be equated to $i$.

As in the case of the other algebras encountered above, we need to
check that the $\Gamma_i$'s obey the commutation relations of the
maximally compact subalgebra of $G_2$, i.e., obey the same
commutation relations as the $k_i$'s, given in appendix
\ref{appendixg2}. We find that the commutation relations are
indeed fulfilled provided we take $\a = i a$, with $a$ solution of
the quadratic equation $16 a^2 + \frac{8}{\sqrt{3}} a - 1 = 0$,
which implies $ \a = - i \frac{\sqrt{3}}{4}$ or $\a = \frac{i}{4
\sqrt{3}}$.  The two different solutions correspond to a non
trivial representation for either the left or the right factor
$SU(2)$. Thus, we see again that the fermions are compatible with
$G_2$-invariance and we are led to introduce the covariant Dirac
operator \be \gamma^\mu D_\mu \psi = \gamma^\mu (\partial_{\m} -
{1 \over 4} \o_{\m,mn} \g^{mn}) \psi - { 1 \over 2}\a \g^{ \m \r
\s } F_{(2) \ \r \s} \psi \ee (with $\a$ equal to one of the above
values) for the spin-$1/2$ field. This is the same expression as
the one that followed from supersymmetry \cite{ChamNic}.

Another approach of this problem is to remember that $G_2$ can be
embedded in $D_4 = SO(4,4)$ \cite{Cremmer:1999du}. The maximal
oxidation is $D=6$ and contains a 2-form in addition to gravity.
After reduction on a circle, we get two dilatons and three
1-forms: the original 2-form and its Hodge dual both reduce to
1-forms, and we have also the Kaluza-Klein 1-form. The model we
are dealing with is obtained by equating these three 1-forms, and
setting the dilatons to zero \cite{Cremmer:1999du}. It is clear
that this projection do respect the covariance of the fermionic
coupling obtained by reduction of (\ref{dnlagf0}). In $D=3$, the
compact gauge group is projected from $SO(4) \times SO(4)$ onto
$SO(4)$, in addition to the unbroken
$SO(1,2)$. All other terms in the connection are indeed set to
zero by the embedding. The $D=6$ spinor can be chosen to have a
definite chirality. Each chirality corresponds to a different
choice of $\a$ after dimensional reduction.

\section{Non-simply laced algebras $B_n$, $C_n$, $F_4$}
\label{nonsimplylaced}

All the non-simply laced algebras can be embedded in simply laced
algebras \cite{Cremmer:1999du}. Therefore, we can find the
appropriate coupling by taking the one obtained for the simply
laced algebras and by performing the same identifications as for
the bosonic sector.

$B_n = SO(n,n+1)$ (with maximal compact subgroup $SO(n) \times
SO(n+1)$) can be obtained from $D_{n+1}$ by modding out the
$\ZZ_2$ symmetry of the diagram. As the $D_{n+1}$ coset can be
oxidised up to $D=n+3$, we must consider a $D=n+3$ Clifford
algebra. The $B_n$ coset has its maximal oxidation in one
dimension lower. If $n$ is even, $D=n+3$ and $D=n+2$ Dirac spinors
are the same: the embedding gives a coupling to a single $D=n+2$
Dirac spinor. Is $n$ is odd, this argument is no longer
sufficient. However, due to the fact that all elements of the
compact subalgebra of $SO(n+1,n+1)$ are represented by a product
of an even number of gamma matrices, we can take a Weyl spinor in
$D=n+3$: it gives a single Dirac spinor in $D=n+2$. It is thus
possible couple the maximal oxidation of the $B_n$ coset to a
single Dirac spinor, such that it reduces to a Dirac coupling in
$D=3$, covariant with respect to $SO(1,2) \times SO(n) \times
SO(n+1)$.  We leave the details to the reader.

For $C_n = Sp(n)$ (with maximal compact subgroup $U(n)$),
the maximal oxidation lives in $D=4$. The embedding in
$A_{2n-1}$ couples the bosonic degrees of freedom to a $D=2n+2$
spinor. As it is an even dimension, the Weyl condition can be
again imposed, so that we get a $D=2n$ Dirac spinor. It is not
possible to reduce further the number of components: the
representation involves product of odd numbers of gamma matrices.
In $D=4$, this gives a coupling to $2^{n-2}$ Dirac spinors.

The situation for $F_4$ is similar, when considering the embedding
in $E_6$. The $E_6$ coset can be oxidised up to $D=8$, with a
consistent fermionic coupling to a Dirac spinor. As the coupling
to the 3-form involves the product of 3 gamma matrices, it is not
possible to impose the Weyl condition. The maximal oxidation of
the $F_4/(SU(2) \times Sp(3))$ coset,
which lives in dimension 6, is thus coupled to a
pair of Dirac spinors.

For $C_n$ and $F_4$, the embeddings just described give a coupling
to respectively $2^{n-2}$ and $2$ Dirac spinors in the maximally
oxidised theory. We have not investigated in detail whether one
could construct invariant theories with a smaller number of
spinors.

\section{$G^{++}$ Symmetry}
\label{G++} The somewhat magic emergence of unexpected symmetries
in the dimensional reduction of gravitational theories has raised
the question of whether these symmetries, described by the algebra
$G$ in three dimensions, are present prior to reduction or are
instead related to toroidal compactification. It has been argued
recently that the symmetries are, in fact, already present in the
maximally oxidized version of the theory (see \cite{dWN1} for
early work on the $E_8$-case) and are part of a much bigger,
infinite-dimensional symmetry, which could be the overextended
algebra $G^{++}$ \cite{JuliaInf,Nicolai,DH1}, the very extended
algebra $G^{+++}$ \cite{West,EH}, or a Borcherds superalgebra
related to it \cite{Henry-LabordereJP}. There are different
indications that this should be the case, including a study of the
BKL limit of the dynamics \cite{BKL}, which leads to
``cosmological billiards" \cite{DHN2}.

In \cite{DHN}, an attempt was made to make the symmetry manifest
in the maximal oxidation dimension by reformulating the system as
a $(1+0)$-non linear sigma model $G^{++}/K(G^{++})$.  The explicit
case of $E_{10}$ was considered.  It was shown that at low levels,
the equations of motion of the bosonic sector of 11-dimensional
supergravity can be mapped on the equations of motion of the non
linear sigma model $E_{10}/ K(E_{10})$.  The matching works for
fields associated with roots of $E_{10}$ whose height does not
exceed 30  (see also \cite{DN2004}).

We now show that this matching works also for Dirac spinors.  We
consider again the explicit case of $E_{10}$ for definiteness.  We
show that the Dirac Lagrangian for a Dirac spinor in eleven
dimensions, coupled to the supergravity three-form as in section
\ref{e8}, is covariant under $K(E_{10})$, at least up to the level
where the bosonic matching is successful.  [For related work on
including fermions in these infinite-dimensional algebras, see
\cite{NicCompact}.]

Our starting point is the action for the non linear sigma model
$E_{10}/ K(E_{10})$ in $1+0$ dimension coupled to Dirac fermions
transforming in a representation of $K(E_{10})$.  We follow the
notations of \cite{DHN2}.  The Lagrangian reads \be \cL =
\frac{1}{2} n^{-1} <\cP \vert \cP> + i \Psi^{\dagger} D_t \Psi
\label{basic1+0}\ee where we have introduced a lapse function $n$
to take into account reparametrization invariance in time. The
$K(E_{10})$ connection is \be \cQ = \sum_{\a \in \Delta_+}
\sum_{s=1}^{mult(\a)} \cQ_{\a,s} K_{\a, s} \ee while the covariant
derivative is \be  D_t \Psi = \dot{\Psi} - \sum_{\a,s}\cQ_{\a,s}
T_{\a,s} \Psi \ee where the $T_{\a,s}$ are the generators of the
representation in which $\Psi$ transforms (there is an infinity of
components for $\Psi$).

In the Borel gauge, the fermionic part of the Lagrangian becomes
\be i \Psi^{\dagger} \dot{\Psi} - \frac{i}{2} \sum_{\a,s}
e^{\a(\b)} j_{\a,s} \Psi^{\dagger} T_{\a,s} \Psi
\label{sigmafermionic}\ee where $\beta^\m$ are now the Cartan
subalgebra variables (i.e., we parametrize the elements of the
Cartan subgoup as $\exp (\b^\m h_\m$)) and $\a(\b)$ the positives
roots. The ``currents" $j_{\a,s}$ (denoted by $\cF_{\a,s}$ in
previous sections) are, as before, the coefficients of the
positive generators in the expansion of the algebra element
$\dot{\cV} \cV^{-1}$, \be \dot{\cV} \cV^{-1} = \dot{\b}^\m h_\m +
\sum_{\a \in \Delta_+} \sum_{s=1}^{mult(\a)} \exp
{\left(\a(\b)\right)} j_{\a,s} E_{\a,s} \ee  We must compare
(\ref{sigmafermionic}) with the Dirac Lagrangian in 11 dimensions
with coupling to the 3-form requested by $E_8$ invariance, \be e
\bpsi \G(\gamma^\mu \p_\mu - \frac{1}{4} \omega_{\mu \ ab}
\gamma^\mu \gamma^{ab} - \frac{1}{2.4!} F_{\mu\nu\rho\sigma}
\gamma^{\mu\nu\rho\sigma} \D) \psi \label{Dirac11}\ee  where $e$
is now the determinant of the space-time vielbein. To make the
comparison easier, we first take the lapse $n$ equal to one
(standard lapse $N$ equal to $e^{-1}$) since both
(\ref{sigmafermionic}) and (\ref{Dirac11}) are reparametrization
invariant in time.  We further split the Dirac Lagrangian
(\ref{Dirac11}) into space and time using a zero shift ($N^k = 0$)
and taking the so-called time gauge for the vielbeins $e^a_\m$,
namely no mixed space-time component. This yields \bea && i
\chi^{\dagger} \left( \dot{\chi} - \frac{1}{4} \omega_{ab}^{R}
\gamma^{ab} \chi - \frac{1}{2.3!} F_{0abc}\gamma^{abc} \chi -
\frac{e}{2 .4! \, 6!}\varepsilon_{abcdp_1
p_2 \cdots p_6} F^{abcd} \gamma^{p_1 \cdots p_6}\chi \right)\nn \\
&& \hspace{1cm} + i \chi^{\dagger} \left(- \frac{e}{2.2!\, 8!}
\omega_{k}^{\; \; ab} \varepsilon_{ab p_1 \cdots p_8} \gamma^k
\gamma^{p_1 \cdots p_8} \chi  + \frac{e}{10!} \varepsilon_{p_1
\cdots p_{10}}\gamma^k \gamma^{p_1 \cdots p_{10}}
\partial_k \chi \right) \label{keykey}\eea where $e$ is now the determinant of
the spatial vielbein and where the Dirac field is taken to be
Majorana (although this is not crucial) and has been rescaled as
$\chi = e^{1/2} \psi$. In (\ref{keykey}), the term
$\omega_{ab}^{R}$ stands for \be \omega_{ab}^{R} = -{1 \over
2}(e_a{}^\m \dot{e}_{\m b} - e_b{}^\m \dot{e}_{\m a}) \ee

A major difference between (\ref{sigmafermionic}) and
(\ref{keykey}) is that $\Psi$ has an infinite number of components
while $\chi$ has only $32$ components. But $\Psi$ depends only on
$t$, while $\chi$ is a spacetime field.  We shall thus assume, in
the spirit of \cite{DHN}, that $\Psi$ collects the values of
$\chi$ and its successive spatial derivatives at a given spatial
point, \be \Psi^{\dagger} = (\chi^{\dagger}, \partial_k
\chi^{\dagger} , \cdots) \ee [The dictionary between $\Psi$ on the
one hand and $\chi$ and its successive derivatives on the other
hand might be more involved (the derivatives might have to be
taken in privileged frames and augmented by appropriate
corrections) but this will not be of direct concern for us here.
We shall loosely refer hereafter to the ``spatial derivatives of
$\chi$" for the appropriate required modifications.] We are thus
making the strong assumption that by collecting $\chi$ and its
derivatives in a single infinite dimensional object, one gets a
representation of $K(E_{10})$.  It is of course intricate to check
this assertion, partly because $K(E_{10})$ is poorly understood
\cite{NicCompact}. Our only justification is that it makes sense
at low levels.

Indeed, by using the bosonic, low level, dictionary of \cite{DHN},
we do see the correct connection terms appearing in (\ref{keykey})
at levels $0$ ($\omega_{ab}^{R}$ term), $1$ (electric field term)
and $2$ (magnetic field term).  The corresponding generators
$\gamma^{a_1 a_2}$, $\gamma^{a_1 a_2 a_3}$ and $\gamma^{a_1 \cdots
a_6}$ do reproduce the low level commutation relations of
$K(E_{10})$.

The matching between the supergravity bosonic equations of motion
and the nonlinear sigma model equations of motion described in
\cite{DHN} goes slightly beyond level $2$ and works also for some
roots at level $3$.  We shall refer to this as ``level $3^-$". To
gain insight into the matching at level $3^-$ for the fermions, we
proceed as in \cite{DHN} and consider the equations in the
homogeneous context of Bianchi cosmologies \cite{DHHS,MHBianchi}
(see also \cite{Saha}). The derivative term $\partial_k \chi$ then
drops out --- we shall have anyway nothing to say about it here,
where we want to focus on the spin connection term $\omega_k^{\;
\; ab}$.  In the homogeneous context, the spin connection term
becomes \be \omega_{abc} = \frac{1}{2}\left(C_{cab} + C_{bca} -
C_{abc} \right)\ee in terms of the structure constants $C^a_{\; \;
bc} = - C^a_{\; \; cb}$ of the homogeneity group expressed in
homogeneous orthonormal frames (the $C^a_{\; \; bc}$ may depend on
time). We assume that the traces $C^a_{\; \; ac}$ vanish since
these correspond to higher height and go beyond the matching of
\cite{DHN}, i.e., beyond level $3^-$. In that case, one may
replace $\omega_{abc}$ by $(1/2)C_{abc}$ in (\ref{keykey}) as can
be seen by using the relation
$$\varepsilon_{abp_1 \cdots p_8}\gamma^k \gamma^{p_1 \cdots p_8} =
\varepsilon_{abp_1 \cdots p_8} \gamma^{k p_1 \cdots p_8} +
\varepsilon_{abkp_2 \cdots p_8} \gamma^{p_2 \cdots p_8}. $$ The
first term drops from (\ref{keykey}) because $\omega^a_{\; \; ba} =
0$, while the second term is completely antisymmetric in $a$, $b$,
$k$. Once $\omega_{abc}$ is replaced by $(1/2)C_{abc}$, one sees
that the remaining connection terms in (\ref{keykey}), i.e., the one
involving a product of nine $\gamma$-matrices, agree with the
dictionary of \cite{DHN}. {}Furthermore, the corresponding level
three generators $\gamma^c \gamma^{p_1 \cdots p_8}$ also fulfill the
correct commutation relations of $K(E_{10})$ up to the requested
order.

To a large extent, the $E_{10}$ compatibility of the Dirac
fermions up to level $3^-$ exhibited here is not too surprising,
since it can be viewed as a consequence of $SL_{10}$ covariance
(which is manifest) and the hidden $E_8$ symmetry, which has been
exhibited in previous sections. The real challenge is to go beyond
level $3^-$ and see the higher positive roots emerge on the
supergravity side. These higher roots might be connected, in fact,
to quantum corrections \cite{DN2005} or higher spin degrees of
freedom.

\section{BKL limit}
\label{BBKKLL} We investigate in this final section how the Dirac
field modifies the BKL behaviour.  To that end, we first rewrite
the Lagrangian (\ref{basic1+0}) in Hamiltonian form.  The
fermionic part of the Lagrangian is already in first order form
(with $i \Psi^{\dagger}$ conjugate to $\Psi$), so we only need to
focus on the bosonic part. The conjugate momenta to the Cartan
fields $\b^\m$ are unchanged in the presence of the fermions since
the time derivatives $\dot{\b}^\m$ do not appear in the connection
$\cQ_{\a,s}$. However, the conjugate momenta to the off-diagonal
variables parameterizing the coset do get modified. How this
affects the Hamiltonian is easy to work out because the time
derivatives of these off-diagonal group variables occur linearly
in the Dirac Lagrangian, so the mere effect of the Dirac term is
to shift their original momenta.  Explicitly, in terms of the
(non-canonical) momentum-like variables \be \Pi_{\a,s} =
\frac{\delta \cL}{\delta j_{\a,s}} \ee introduced in
\cite{MatNic,DHN2}, one finds \be \Pi_{\a,s} = \Pi_{\a,s}^{old}  -
\frac{1}{2} \exp \left( \a(\b) \right) J^F_{\a,s} \ee where
$\Pi_{\a,s}^{old}$ is the bosonic contribution (in the absence of
fermions) and where $J^F_{\a,s}$ are the components of the
fermionic $K(G^{++})$-current, defined by \be J^F_{\a,s} = i
\Psi^\dagger T_{\a,s} \Psi . \ee The currents $J^F_{\a,s}$ are
real.

It follows that the Hamiltonian associated with (\ref{basic1+0})
takes the form \be \cH = n \left( \frac{1}{2} G^{\m \n} \pi_\m
\pi_\n + \sum_{\a \in \Delta_+} \sum_{s=1}^{mult(\a)} \exp
{\left(-2 \a(\b)\right)} \left(\Pi_{\a,s} + \frac{1}{2} \exp
\left( \a(\b) \right) J^F_{\a,s} \right)^2 \right)
\label{Hamiltonian}\ee If one expands the Hamiltonian, one gets
\be \cH = n \left( \frac{1}{2} G^{\m \n} \pi_\m \pi_\n +
\sum_{\a,s} \exp {\left(-2 \a(\b)\right)} \Pi_{\a,s}^2  +
\sum_{\a,s} \exp \left(- \a(\b) \right) \Pi_{\a,s} J^F_{\a,s} +
\frac{1}{4} C \right) \ee where $C$ is (up to a numerical factor)
the quadratic Casimir of the fermionic representation, \be C =
\sum_{\a,s} (J^F_{\a,s})^2 . \ee  We see that, just as in the pure
bosonic case, the exponentials involve only the positive roots
with negative coefficients. However, we obtain, in addition to the
bosonic walls, also their square roots. All the exponentials in
the Hamiltonian are of the form $\exp (-2 \rho(\b))$, where
$\rho(\b)$ are the positive roots or half the positive roots.

In order to investigate the asymptotic BKL limit $\b^\m
\rightarrow \infty$, we shall treat the $K(G^{++})$-currents as
classical real numbers and consider their equations of motion that
follow from the above Hamiltonian, noting that their Poisson
brackets $[J^F_{\a,s}, J^F_{\a',s'}]$ reproduce the
$K(G^{++})$-algebra. This is possible because the Hamiltonian in
the Borel gauge involves only the $\Psi$-field through the
currents. This is a rather remarkable property. [A ``classical"
treatment of fermions is well known to be rather delicate.  One
can regard the dynamical variables, bosonic and fermionic, as
living in a Grassmann algebra.  In that case, bilinear in fermions
are ``pure souls" and do not influence the behaviour of the
``body" parts of the group elements, which are thus trivially
governed by the same equations of motion as in the absence of
fermions. However, it is reasonable to expect that the currents
$J^F_{\a,s}$ have a non trivial classical limit (they may develop
non-vanishing expectation values) and one might treat them
therefore as non-vanishing real numbers.  This is technically
simple here because the currents obey closed equations of motion.
It leads to interesting consequences.]

Next we observe that $[J^F_{\a,s}, C]= 0$. It follows that the
quadratic Casimir $C$ of the fermionic representation is
conserved.  Furthermore, it does not contribute to the dynamical
Hamiltonian equations of motion for the group variables or the
currents. By the same reasoning as in \cite{DHN2}, one can then
argue that the exponentials tend to infinite step theta functions
and that all variables except the Cartan ones, i.e., the
off-diagonal group variables and the fermionic currents,
asymptotically freeze in the BKL limit.

Thus, we get the same billiard picture as in the bosonic case, with
same linear forms characterizing the walls (some of the exponential
walls are the square roots of the bosonic walls). But the free
motion is governed now by the Hamiltonian constraint \be G^{\m \n}
\pi_\m \pi_\n + M^2 \approx 0\ee  with $M^2 = C/2>0$.   This implies
that the motion of the billiard ball is timelike instead of being
lightlike as in the pure bosonic case. This leads to a non-chaotic
behaviour, even in those cases where the bosonic theory is chaotic.
Indeed, a timelike motion can miss the walls, even in the hyperbolic
case. This is in perfect agreement with the results found in
\cite{BK} for the four-dimensional theory.

Our analysis has been carried out in the context of the sigma model
formulation, which is equivalent to the Einstein-Dirac-p-form system
only for low Kac-Moody levels. However, the low levels roots are
precisely the only relevant ones in the BKL limit (``dominant
walls").  Thus, the analysis applies also in that case.  Note that
the spin $1/2$ field itself does not freeze in the BKL limit, even
after rescaling by the quartic root of the determinant of the
spatial metric, but asymptotically undergoes instead a constant
rotation in the compact subgroup, in the gauge $n=1$ (together with
the Borel gauge). Note also that the same behaviour holds if one
adds a mass term to the Dirac Lagrangian, since this term is
negligible in the BKL limit, being multiplied by $e$, which goes to
zero.

One might worry that the coefficients $\Pi_{\a,s} J^F_{\a,s}$ of the
square roots of the bosonic walls have no definite sign. This is
indeed true but generically of no concern for the following reason:
in the region $\a(\b) <0$ outside the billiard table where the
exponential terms are felt and in fact blow up with time at a given
configuration point ($\a(\b)\rightarrow - \infty$) \cite{DHN2}, the
wall $\exp (- 2 \a(\b))$ dominates the wall $\exp (- \a(\b))$ coming
from the fermion and the total contribution is thus positive. The
ball is repelled towards the billiard table.

\section{Conclusions}
In this paper, we have shown that the Dirac field is compatible
with the hidden symmetries that emerge upon toroidal dimensional
reduction to three dimensions, provided one appropriately fix its
Pauli couplings to the $p$-forms.  We have considered only the
split real form for the symmetry (duality) group in three
dimensions, but similar conclusions appear to apply to the
non-split forms (we have verified it for the four-dimensional
Einstein-Maxwell-Dirac system, which leads to the $SU(2,1)/S(U(2)
\times U(1))$ coset in three dimensions).  We have also indicated
that the symmetry considerations reproduce some well known
features of supersymmetry when supersymmetry is available.

We have also investigated the compatibility of the Dirac field
with the conjectured infinite-dimensional symmetry $G^{++}$ and
found perfect matching with the non-linear sigma model equations
minimally coupled to a $(1+0)$ Dirac field, up to the levels where
the bosonic matching works.

{}Finally, we have argued that the Dirac fermions destroy chaos
(when it is present in the bosonic theory), in agreement with the
findings of \cite{BK}.  This has a rather direct group theoretical
interpretation (motion in Cartan subalgebra becomes timelike) and
might have important implications for the pre-big-bang
cosmological scenario and the dynamical crossing of a cosmological
singularity \cite{Veneziano,al}.

It would be of interest to extend these results to include the
spin $3/2$ fields, in the supersymmetric context. In particular,
11-dimensional supergravity should be treated.  To the extent that
$E_{10}$ invariance up to the level $3^-$ is a mere consequence
of $E_8$ invariance in three dimensions and $SL_{10}$ convariance,
one expects no new feature in that respect since the reduction to
three dimensions of full supergravity is indeed known to be $E_8$
invariant \cite{Marcus:1983hb}.  But perhaps additional structure
would emerge.  Understanding the BKL limit might be more
challenging since the spin $3/2$ fields might not freeze, even
after rescaling.

\subsection*{Aknowledgments}
We are grateful to Hermann Nicolai for a useful discussion and to Arjan
Keurentjes for a helpful remark.  This
work is partially supported by IISN - Belgium (convention
4.4505.86), by the ``Interuniversity Attraction Poles Programme --
Belgian Science Policy'' and by the European Commission FP6
programme MRTN-CT-2004-005104, in which we are associated to
V.U.Brussel.

\appendix

\section{Conventions for $D_n$}
\label{appendixdn}

\subsection{Generators and Algebra}
We can express the positive generators of $D_{n}$ as follow, \bea
e_{b_{ij}} & \equiv & [e_i,[...,[e_{j-2},e_{j-1}]...] \ \ i<j \no
\\ e_{b_i} & \equiv & [\tilde{e}_{n-1},e_{b_{1i}}] \no
\\
e_{a_i} & \equiv & [e_{b_{in-1}},e_{n-1}] \\
e_{a_{ij}} & \equiv & [[e_n,e_{b_{2j}}],e_{b_{1i}}] \ i<j \eea
where $\tilde{e}_{n-1} = [e_{n},[e_{b_{2n-1}},e_{n-1}]]$ and where
$ i = 1,... ,n-1$ and $e_{b_{ii}}$ must be understood as being
absent. The Chevalley-Serre generators of $D_n$, namely $\{e_m \
\arrowvert \ m = 1,...,n \}$, are given by $e_i = e_{b_{i i+1}}$
($i=1,...,n-2$), $e_{n-1} = e_{a_{n-1}}$ and $e_n = e_{a_{12}}$.
These generators are associated to the vertices numbered as shown
in the following Dynkin diagram,
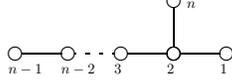
\begin{center}
\scalebox{.5}{
\begin{picture}(180,60)
\put(5,-5){$n-1$} \put(45,-5){$n-2$} \put(85,-5){$3$}
\put(125,-5){$2$} \put(165,-5){$1$} \put(140,45){$n$}
\thicklines \multiput(10,10)(40,0){4}{\circle{10}}
\multiput(95,10)(40,0){2}{\line(1,0){30}}
\dashline[0]{2}(55,10)(65,10)(75,10)(85,10)
\put(130,50){\circle{10}} \put(130,15){\line(0,1){30}}
\multiput(130,10)(40,0){2}{\circle{10}}
\put(15,10){\line(1,0){30}}
\end{picture}
} \end{center} Their non vanishing commutation relations are \bea
\ [e_{b_{ij}},e_{b_{mn}}] &=& \d_{jm} e_{b_{in}} - \d_{in}e_{b_{mj}} \no \\
\ [e_{b_{ij}},e_{b_m} ] &=& - \d_{im} e_{b_j}\no \\
\ [e_{b_{ij}},e_{a_{mn}}] &=& -\d_{im} e_{a_{jn}} - \d_{in}e_{a_{mj}} + \d_{im} e_{a_{nj}} \no\\
\ [e_{b_{ij}},e_{a_m}] &=& \d_{jm} e_{a_i} \no \\
\ [e_{a_{ij}},e_{a_m} ]& = &- \d_{im} e_{b_j} + \d_{jm} e_{b_i}
\eea Notations are similar for the negative generators (with $f$'s
instead of $e$'s). One easily verifies that the normalization
factors $N_\a$ are all equal to one, $K(e_\a, f_\b) = - \delta_{\a
\b}$.

\subsection{Compact subgroup}
As explained above, the involution $\tau$ is such that $\tau(h_i)=
-h_i$, $\tau(e_{\a})= f_{\a}$ and $\tau(f_{\a})=e_{\a}$ so that a
basis of the maximally compact subalgebra of $D_n$ reads $k_{\a} =
e_{\a} + f_{\a}$ where $ \a = \{a_{ij},a_i,b_{ij},b_i \}$ and
$i<j= 1,..., n-1$. The commutation relations of the $k_\a$'s are
\bea \ [k_{b_{ij}},k_{b_{mn}}] &=& \d_{jm} k_{b_{in}} -
\d_{in}k_{b_{mj}} +
\d_{im}(k_{b_{nj}}-k_{b_{jn}}) + \d_{jn}(k_{b_{mi}} - k_{b_{im}}) \no \\
\ [k_{b_{ij}},k_{b_m} ] &=& - \d_{im} k_{b_j} + \d_{jm}k_{b_{i}} \no \\
\ [k_{b_{ij}},k_{a_{mn}}] &=& -\d_{im} k_{a_{jn}} -
\d_{in}k_{a_{mj}} +
\d_{im} k_{a_{nj}} + \d_{jm}k_{b_{in}} + \d_{jn}(k_{b_{mi}}-k_{b_{im}}) \no\\
\ [k_{b_{ij}},k_{a_m}] &=& \d_{jm} k_{a_i} - \d_{im} k_{a_{j}} \no \\
\ [k_{b_{ij}},k_{a_m} ]& = &- \d_{im} k_{a_j} + \d_{jm} k_{a_i}\no \\
\ [k_{b_i},k_{b_j}] &=& -k_{b_{ij}}+k_{b_{ji}} \no \\
\ [k_{b_{i}},k_{a_{mn}} ]& = &- \d_{in} k_{a_m} + \d_{im} k_{a_n}
\no \\
\ [k_{b_i},k_{a_j}] &=& -k_{a_{ij}}+k_{a_{ji}} \no \\
\ [k_{a_{ij}},k_{a_{mn}}] &=& \d_{jm} k_{b_{in}} -
\d_{in}k_{b_{mj}} +
\d_{im}(k_{b_{nj}}-k_{b_{jn}}) + \d_{jn}(k_{b_{mi}} - k_{b_{im}}) \no \\
\ [k_{a_{ij}},k_{a_m} ]& = &- \d_{im} k_{b_j} + \d_{jm} k_{b_i}
\eea By going to the new basis $\{k_b + k_a, k_b - k_a \}$, one
easily recognizes the algebra $so(n) \oplus so(n)$.

\subsection{Embedding of $A_{n-1}$}
The gravitational subalgebra $A_{n-1}$ is generated by $h_1,
\cdots, h_{n-2}, \tilde{h}_{n-1}$ (Cartan generators), $e_1,
\cdots, e_{n-2}, \tilde{e}_{n-1}$ (raising operators) and $f_1,
\cdots, f_{n-2}, \tilde{f}_{n-1}$ (lowering operators), with
$\tilde{h}_{n-1} = - h_n - h_2 - h_3 - \cdots - h_{n-1}$. The
simple root $\tilde{\a}_{n-1}$ is connected to $\a_1$ only, with a
single link. Note that although it is a simple root for the
gravitational subalgebra $A_{n-1}$, it is in fact the highest root
of the $A_{n-1}$ subalgebra associated with the Dynkin subdiagram
$n, 2, 3 , \cdots, n-1$.

\section{$E_8$ algebra}
\label{E8}

We take a basis of the Cartan subalgebra $(h_i)$, such that \be
\begin{array}{rcl}
\G[ f_{ij} , e_{ij} \D] &=& h_i - h_j \\
\G[ f_{i} , e_{i} \D] &=& -h_i \\
\G[ \tilde{f}_{ijk} , \te_{ijk} \D] &=& \frac{1}{3} (h_1 + h_2 +
\ldots +
h_8) -h_i - h_j -h_k \\
\G[ \tilde{f}_{ij} , \te_{ij} \D] &=& - \frac{1}{3} (h_1 + h_2 +
\ldots + h_8) + h_i + h_j
\end{array}
\ee \be
\begin{array}{rclrcl}
\G[ h_i , e_{ij} \D] &=& e_{ij} & \qquad \G[ h_i , f_{ij} \D] &=& -f_{ij} \\
\G[ h_j , e_{ij} \D] &=& -e_{ij} & \G[ h_j , f_{ij} \D] &=& f_{ij} \\
\G[ h_i , e_i \D] &=& - e_i & \G[ h_i , f_i \D] &=& f_i \\
\G[ h_i , \te_{ijk} \D] &=& -\te_{ijk} & \G[ h_i , \tf_{ijk} \D]
&=&
\tf_{ijk} \\
\G[ h_i , \te_{jk} \D] &=& -\te_{jk} & \G[ h_i , \tf_{jk} \D] &=&
\tf_{jk}
\end{array}
\ee where distinct indices are supposed to have different values.
The vectors associated with the simple roots are $e_{i\,i+1}$,
$\te_{123}$.

Other non vanishing commutations relations are the following, with
the same convention on indices. \be
\begin{array}{rclrcl}
\G[ e_{ij} , e_{jk} \D] &=& e_{ik}
& \qquad \G[ f_{ij} , f_{jk} \D] &=& f_{ik} \\
\G[ \te_{ijk}, e_{kl} \D] &=& e_{ijl}
& \G[ \tf_{ijk}, f_{kl} \D] &=& f_{ijl} \\
\G[ \te_{ijk} , \te_{lmn} \D] &=& \frac{1}{2} \epsilon^{ijklmnpq}
\te_{pq} & \G[ \tf_{ijk} , \tf_{lmn} \D] &=& \frac{1}{2}
\epsilon^{ijklmnpq} \tf_{pq}
\\
\G[ e_{ij} , \te_{jk} \D] &=& \te_{ik}
& \G[ f_{ij} , \tf_{jk} \D] &=& \tf_{ik} \\
\G[ \te_{ijk} , \te_{jk} \D] &=& e_i
& \G[ \tf_{ijk} , \tf_{jk} \D] &=& f_i \\
\G[ e_i , e_{ij} \D] &=& e_j & \G[ f_i , f_{ij} \D] &=& f_j
\end{array}
\ee \be
\begin{array}{rcllrcll}
\G[ f_{ij} , e_{kj} \D] &=& e_{ki} & \textrm{ if } i>k & \qquad
\G[ f_{ij} , e_{kj} \D] &=& -f_{ik} & \textrm{ if } i<k \\
\G[ f_{ji} , e_{jk} \D] &=& -e_{ik} & \textrm{ if } i<k &
\G[ f_{ji} , e_{jk} \D] &=& f_{ki} & \textrm{ if } i>k \\
\G[ f_{ij} , \te_{klj} \D] &=& e_{kli} &&
\G[ e_{ij} , \tf_{klj} \D] &=& f_{kli} \\
\G[ \tf_{ijk} , \te_{ijl} \D] &=& -e_{kl} & \textrm{ if } k<l &
\G[ \tf_{ijk} , \te_{ijl} \D] &=& f_{lk} & \textrm{ if } k>l \\
\G[ f_{ji} , \te_{jk} \D] &=& -\te_{ik} &&
\G[ e_{ji} , \tf_{jk} \D] &=& -\tf_{ik} \\
\G[ \tf_{ijk} , \te_{lm} \D] &=& \multicolumn{2}{l}{-\frac{1}{3!}
\epsilon^{ijklmnpq} \te_{npq}} & \G[ \te_{ijk} , \tf_{lm} \D] &=&
\multicolumn{2}{l}{-\frac{1}{3!} \epsilon^{ijklmnpq} \tf_{npq}}
\\
\G[ f_{ij} , e_j \D] &=& e_i &&
\G[ e_{ij} , f_j \D] &=& f_i \\
\G[ \tf_{ij} , \te_{ik} \D] &=& e_{kj} & \textrm{ if } j>k &
\G[ \tf_{ij} , \te_{ik} \D] &=& -f_{jk} & \textrm{ if }\ j<k \\
\G[ \tf_{ijk} , e_k \D] &=& -\te_{ij} &&
\G[ \te_{ijk} , f_k \D] &=& -\tf_{ij} \\
\G[ \tf_{ij} , e_k \D] &=& \te_{ijk} &&
\G[ \te_{ij} , f_k \D] &=& \tf_{ijk} \\
\G[ f_{i} , e_j \D] &=& -e_{ij} & \textrm{ if } i<j & \G[ f_{i} ,
e_j \D] &=& f_{ji} & \textrm{ if } i>j
\end{array}
\ee

The Chevalley-Serre generators are $h_i-h_{i+1}$, $h_{123} \equiv
\frac{1}{3}(h_1 + \cdots + \h_8) - h_1 - h_2 - h_3$, $e_{i i+1}$,
$\te_{123}$,  $f_{i i+1}$ and $\tf_{123}$.  The scalar products of
the $h_i$'s are $K(h_i, h_j) = \delta_{ij} + 1$ and the factors
$N_{\alpha}$ are equal to unity.

The generators of the compact subalgebra $\mathrm{so}(16)$ are \be
\begin{array}{rcll}
k_{ij} &=& e_{ij} + f_{ij} & \mathrm{for} \ i<j \no\\
k_i &=& e_i+f_i \no\\
\tk_{ijk} &=& \te_{ijk} + \tf_{ijk} \no\\
\tk_{ij} &=& \te_{ij} + \tf_{ij} \rlap{\ .}
\end{array}
\ee It is convenient to define $k_{ij} = - k_{ji} = -e_{ji} -
f_{ji}$ for $i>j$. Their non vanishing commutators are \be
\begin{array}{rclrcl}
\G[ k_{ij} , k_{jk}\D] &=& k_{ik} &
\G[ \tk_{ijk} , k_{kl}\D] &=& \tk_{ijl} \\
\G[ \tk_{ijk} , \tk_{lmn}\D] &=& \frac{1}{2} \epsilon^{ijklmnpq}
\tk_{pq} &
\G[ \tk_{ijk} , \tk_{ijl}\D] &=& - k_{kl} \\
\G[ k_{ij} , \tk_{jk}\D] &=& \tk_{ik} &
\G[ \tk_{ijk} , \tk_{jk}\D] &=& k_{l} \\
\G[ \tk_{ijk} , \tk_{lm}\D] &=& - \frac{1}{3!} \epsilon^{ijklmnpq}
\tk_{npq} &
\G[ \tk_{ij} , \tk_{ik}\D] &=& - k_{jk} \\
\G[ k_{ij} , k_j \D] &=& k_i &
\G[ \tk_{ijk} , k_k \D] &=& -\tk_{ij} \\
\G[ \tk_{ij} , k_k \D] &=& \tk_{ijk} & \G[ k_{i} , k_j \D] &=&
-k_{ij}
\end{array}
\label{so16} \ee where it is assumed that distinct indices have
different values.

\section{Algebra $G_{2(2)}$}
\label{appendixg2}

Let $e_1$ and $e_2$ be the positive Chevalley generators of $G_2$
corresponding to the two simple roots $\a_1$ and $\a_2$. The other
positive generators are \bea
e_3 &=& [e_2,e_1] \hspace{1cm} e_4 = [e_2,[e_2,e_1]] \nn \\
e_5 &=& [e_2,[e_2,[e_2,e_1]]] \hspace{1cm} e_6 =
[[e_2,[e_2,[e_2,e_1]]], e_1] .\eea Their non vanishing commutation
relations are,
\bea \ [e_1, e_2 ] &=& -e_3 \hspace{1cm} [e_1, e_5 ] = - e_6 \nn \\
\ [e_2, e_3 ] &=& e_4 \hspace{1cm} [e_2, e_4 ] = e_5 \nn \\
\ [e_3,e_4] &=& e_6\hspace{1cm} \eea

The normalizing factors $N_\a$ for the simple roots are given by
$N_1 = 1$ and $N_2 = 3$ since $(\a_1 \vert \a_1) = 2$ and $(\a_2
\vert \a_2) = \frac{2}{3}$.  It follows that $N_3 = 3$, $N_4 =
12$, $N_5 = 36$ and $N_6 = 36$.  We define the vectors $\e_i$ in
order to absorb these factors, i.e., $\e_1 = e_1$, $\e_2 =
\frac{1}{\sqrt{3}} e_2$, $\e_3 = \frac{1}{\sqrt{3}} e_3$, $\e_4 =
\frac{1}{2\sqrt{3}} e_4$, $\e_5 = \frac{1}{6} e_5$, $\e_6 =
\frac{1}{6} e_6$. This implies $K(\e_i, \tau(\e_i)) = -1$.

We take as compact generators $k_i = \e_i + \tau(\e_i)$.  The
commutators of the compact subalgebra are \bea && [k_1, k_2] = -
k_3, \qquad [k_1, k_3] = k_2, \qquad [k_1, k_4] = 0, \nonumber \\
&& [k_1, k_5] = - k_6, \qquad [k_1, k_6] = k_5 , \qquad [k_2, k_3]
= \frac{2}{\sqrt{3}} k_4 - k_1 \nonumber \\ && [k_2, k_4] = k_5 -
\frac{2}{\sqrt{3}} k_3 , \qquad [k_2, k_5] = - k_4 , \qquad [k_2,
k_6] = 0, \nonumber \\ &&  [k_3, k_4] =  k_6 + \frac{2}{\sqrt{3}}
k_2 , \qquad [k_3, k_5] =  0 , \qquad [k_3, k_6] = - k_4 ,
\nonumber
\\ && [k_4, k_5] = k_2 , \qquad [k_4, k_6] =  k_3, \qquad [k_5,
k_6] = - k_1. \eea  In the basis \bea && \xi_1 = \frac{1}{4}(3 k_1
+ \sqrt{3} k_4),\qquad \xi_2 = \frac{1}{4}(\sqrt{3} k_2 - 3
k_6),\qquad \xi_3 = - \frac{1}{4} (\sqrt{3} k_3 + 3 k_5) \nn \\ &&
X_1 = \frac{1}{4}(k_1 - \sqrt{3} k_4),\qquad X_2 =
\frac{1}{4}(\sqrt{3} k_2 + k_6),\qquad X_3 = - \frac{1}{4}
(\sqrt{3} k_3 - k_5),\eea the commutation relations read \be
[\xi_i, \xi_j] = \varepsilon_{ijk} \xi_k, \qquad [\xi_i, X_j] = 0,
\qquad [X_i, X_j] = \varepsilon_{ijk} X_k \ee and reveal the
$su(2) \oplus su(2)$ structure of the algebra.

\end{document}